\documentclass[11pt,preprint]{aastex}
\usepackage{times}
\usepackage{amssymb}
\usepackage{amsmath}
\usepackage{xspace}
\usepackage{natbib}

\newcommand{\degg}{\hbox{$^\circ$}\xspace}

\newcommand{\xmm}{\emph{XMM-Newton}\xspace}
\newcommand{\xmmrgs}{\emph{XMM-Newton/RGS}\xspace}
\newcommand{\suzaku}{\emph{Suzaku}\xspace}
\newcommand{\asca}{\emph{ASCA}\xspace}

\newcommand{\chandra}{\emph{Chandra}\xspace}

\newcommand{\arcs}{\hbox{$^{\prime\prime}$}}

\newcommand{\Msun}{\hbox{$\rm\thinspace M_{\odot}$}}
\newcommand{\ls}
{\mathrel{\hbox{\rlap{\hbox{\lower4pt\hbox{$\sim$}}}\hbox{$<$}}}}
\newcommand{\gs}
{\mathrel{\hbox{\rlap{\hbox{\lower4pt\hbox{$\sim$}}}\hbox{$>$}}}}

\newcommand{\logxi}{\log (\xi/\rm{erg\,cm\,s}^{-1})}

\received{}
\begin{document}
\title{High Resolution X-ray Spectroscopy of the Seyfert 1, Mrk\,1040. 
Revealing the Failed Nuclear Wind with Chandra.}
\shorttitle{The Failed Wind of Mrk\,1040}
\shortauthors{Reeves et al.}
\author{J.\,N. Reeves\altaffilmark{1,2}, V. Braito\altaffilmark{1,3}, E. Behar\altaffilmark{4,5}, 
T.\,C. Fischer\altaffilmark{6}, S.\,B. Kraemer\altaffilmark{7}, 
A. Lobban\altaffilmark{2,8}, E. Nardini\altaffilmark{2,9},
D. Porquet\altaffilmark{10}, T.\,J. Turner\altaffilmark{11}}
\altaffiltext{1}{Center for Space Science and Technology, 
University of Maryland 
Baltimore County, 1000 Hilltop Circle, Baltimore, MD 21250, USA; jreeves@umbc.edu}
\altaffiltext{2}{Astrophysics Group, School of Physical and Geographical Sciences, Keele 
University, Keele, Staffordshire, ST5 5BG, UK; j.n.reeves@keele.ac.uk}
\altaffiltext{3}{INAF - Osservatorio Astronomico di Brera, Via Bianchi 46 I-23807 Merate (LC), Italy}
\altaffiltext{4}{Dept of Physics, Technion, Haifa 32000, Israel}
\altaffiltext{5}{Department of Astronomy, University of Maryland, College Park, MD 20742, USA}
\altaffiltext{6}{Astrophysics Science Division, NASA Goddard Space Flight Center, Code 665, Greenbelt, MD 20771, USA}
\altaffiltext{7}{Institute for Astrophysics and Computational Sciences, Department of Physics, The Catholic University of America, Washington, DC 20064, USA}
\altaffiltext{8}{Dept of Physics and Astronomy, University of Leicester, University Road, Leicester LE1 7RH, UK}
\altaffiltext{9}{INAF - Arcetri Astrophysical Observatory, Largo Enrico Fermi 5, I-50125 Firenze, Italy}
\altaffiltext{10}{Universit\'{e} de Strasbourg, CNRS, Observatoire astronomique de Strasbourg, UMR 7550, F-67000 Strasbourg, France}
\altaffiltext{11}{Department of Physics, 
University of Maryland Baltimore County, 1000 Hilltop Circle, Baltimore, MD 21250, USA}

\begin{abstract}
High resolution X-ray spectroscopy of the warm absorber in the nearby X-ray bright Seyfert 1 
galaxy, Mrk 1040 is presented. The observations were carried out in the 2013--2014 
timeframe using the Chandra High Energy Transmission Grating with a total exposure of 200\,ks. A multitude of absorption lines from Ne, Mg and Si are detected from a wide variety of ionization states.
In particular, the detection of inner K-shell absorption lines from Ne, Mg and Si, from charge states 
ranging from F-like to Li-like ions, suggests the presence of a substantial amount of low ionization 
absorbing gas, illuminated by a steep soft X-ray continuum.
The observations reveal at least 3 warm absorbing components ranging in ionization 
parameter from $\logxi = 0-2$ and with column densities of $N_{\rm H} =1.5-4.0 \times 10^{21}$\,cm$^{-2}$.  
The velocity profiles imply that the outflow velocities 
of the absorbing gas are low and within $\pm100$\,km\,s$^{-1}$
of the systemic velocity of Mrk\,1040, which suggests any outflowing gas may have stalled in this AGN 
on large enough scales. The warm absorber is likely located 
far from the black hole, within 300\,pc of the nucleus and is spatially coincident with 
emission from an extended Narrow Line Region as seen in the {\it HST} images.
The iron K band spectrum reveals only narrow emission lines, with Fe K$\alpha$ at 6.4 keV consistent with 
originating from reflection off Compton thick pc-scale reprocessing gas.
\end{abstract}


\section{Introduction}

Material existing within the vicinity of the super-massive black hole in an Active Galactic Nucleus (AGN) 
can significantly modify the resultant X-ray spectrum. 
Consequently, detailed studies of the reprocessed primary X-ray emission can reveal valuable information about the kinematics and geometry 
of the central nucleus. In the soft X-ray band (below 2\,keV), the
dominant reprocessing mechanism is absorption by photoionized (or
 ``warm'') material along the line of sight. In fact 
systematic X-ray studies of AGN with \asca, \chandra, \xmm\ and
\suzaku\ have established that at least half of all type 1 Seyferts
host warm absorbers
\citep{reynolds1997,george1998,crenshaw2003,porquet2004,blustin2005,mckernan2007,tombesi2013}.
When observed at high spectral resolution, 
such as with the X-ray grating spectrometers onboard either \chandra\ or
\xmm, the warm absorber is seen to give rise to numerous narrow
absorption lines from astrophysically abundant elements such as C,
N, O, Ne, Mg, Si, S and Fe \citep{kaastra2000,kaspi2002,blustin2002,mckernan2003}.  
The warm absorption is often seen to be outflowing and can arise
from gas covering a wide (several orders of magnitude)  
range of column densities and ionization parameters. 

Indeed the X-ray spectral signatures of the warm absorber 
range can arise from very low ionization gas from ions with low charge states; 
such as from the Unresolved Transition Array (UTA) of M-shell iron ($<$ Fe\,\textsc{xvii})
at $\sim16$\AA~ \citep{sako2001,behar2001} or from inner (K-shell)
transitions of L-shell ions from lighter elements, such as  
O through to Si \citep{behar2002,gu2005}.  
On the other hand, the absorbing gas can also be very highly ionized, in some cases arising from H-like and He-like iron, 
where such highly ionized gas may originate from an accretion disk wind 
\citep[e.g.,][]{reeves2004a,risaliti2005,braito2007,turner2008,tombesi2010a,gofford2013}.
The associated absorption lines are often blueshifted, thus implying
outflowing winds, with velocities ranging from a few hundred to a few
thousand km\,s$^{-1}$, or even exceeding 10000\,km\,s$^{-1}$ in the
case of the so-called ultra fast outflows
\citep{chartas2002,pounds2003b,reeves2003,tombesi2013,gofford2013,nardini2015,tombesi2015,longinotti2015}.

With the exception of the ultra fast outflows, which are often highly ionized and may originate from an accretion disk wind on sub-pc scales, 
most soft X-ray warm absorbers are thought to be typically located at fairly large distances from the central black hole. 
This is as a result of their low ionization parameters and outflow velocities ($<1000$\,km\,s$^{-1}$), their (relative) lack of variability, plus in some
    cases from their inferred low densities (e.g., NGC\,3783:
    \citealt{behar2003,krongold2005}; Mrk\,279:
    \citealt{Scott2004,Ebrero2010}; NGC\,4051:
    \citealt{steenbrugge2009}; Mrk\,290: \citealt{Zhang2011}; and
    Mkn\,509: \citealt{kaastra2012}). These soft X-ray warm
    absorbers can be associated with, for example, 
  a wind originating from the putative parsec scale torus
  (\citealt{blustin2005}) or the latter stages of an accretion
  disc wind which has propagated out to larger radii
  (\citealt{proga&kallman2004,tombesi2013}), or from 
gas associated with the AGN Narrow Line Regions \citep{crenshaw2000,crenshaw&kraemer2000}. 
Although the exact kinematic contribution of the warm absorbers to the total AGN 
luminosity can be difficult to assess \citep{crenshaw&kraemer2012},
they may still play a key role in shaping the wide scale  
properties of AGN host galaxies \citep{hopkins&elvis2010}. 

\subsection{The Target: Markarian 1040}

Mrk 1040 (also known as NGC\,931) is a bright, nearby ($z = 0.016652$, \citealt{huchra1999}) type-1 Seyfert galaxy. 
The likely black hole mass is $\log(M_{\rm BH}/M_{\rm \odot})=7.64\pm0.40$ \citep{zhou2010,demarco2013}, as estimated from its stellar velocity dispersion \citep{nelson&whittle1995,tremaine2002}. 
In X-rays, Mrk\,1040 was observed by \asca\ in August 1994 \citep{reynolds1995} and the spectral features were interpreted as arising 
from a warm absorber plus a strong broad, fluorescent Fe\,K$\alpha$ emission line 
(FWHM $\sim 16\,000$--$70\,000$\,km\,s$^{-1}$; $EW = 500 \pm 250$\,eV). 
It is also bright in the hard X-ray bandpass, having been detected in the 70 month 
Swift BAT catalogue, with a 14-195\,keV band flux of $6.1\times10^{-11}$\,erg\,cm$^{-2}$\,s$^{-1}$ 
\citep{baumgartner2013}.
Mrk 1040 was subsequently observed by \xmm\ in February 2009 
when the source flux was high, with a total exposure of $\sim$84\,ks. The spectrum revealed all the classic Seyfert 1 
characteristics; a soft X-ray excess, 
a deep warm absorber below 2\,keV and a strong Fe K$\alpha$ line profile \citep{tripathi2011}. The 
AGN is also strongly variable on timescales of $10^{4}$\,s and an
energy dependent lag was claimed from the \xmm\ observation \citep{tripathi2011}, similar to the soft X-ray (or negative) lags that have been claimed in other AGN \citep{demarco2013}. Yet despite its X-ray brightness, similar to other bright 
and nearby Seyfert 1 galaxies, the X-ray spectrum of Mrk\,1040 
is comparatively understudied and no high resolution X-ray spectrum has yet been published on this AGN.

The host galaxy of Mrk\,1040 is a Sbc spiral galaxy and is viewed at high inclination of 
$b/a = 0.21$ \citep{dezotti1985,amram1992}. 
The UV continuum is heavily reddened, which \cite{crenshaw&kraemer2001} 
suggested was due to dust in the plane of the host galaxy. 
Based on the width of H$\beta$, Full Width at Half Maximum (FWHM) of $\sim 1830$\,km\,s$^{-1}$, 
it is classified as a Narrow-Line Seyfert 1 \citep{osterbrock1982}. 
If the narrowness of the Balmer lines is due to a line-of-sight close to the rotation axis of 
the black hole/accretion disk \citep{kraemer2012}, then 
our view of the AGN in Mrk 1040 may be roughly pole-on, which suggests that the AGN is highly 
inclined with respect to the disk of the host galaxy. 

This paper presents the analysis of a series of \chandra\ 
High Energy Transmission Grating (HETG) \citep{weisskopf2000,canizares2005}
 observations of Mrk\,1040 from October 2013 to March 2014, with 
a total exposure of $\sim 200$\,ks. Indeed the observations provide the 
first reported high spectral resolution observations of this Seyfert 1 galaxy over 
the X-ray band. Subsequent observations with \xmm\ were also obtained 
during 2015, with a total exposure of 175\,ks in the RGS; these together with the archival observation from 2009 will be presented in a subsequent paper (hereafter paper II). 
One of the primary goals was to study the warm absorber in this AGN in
unprecedented detail and resolution. Here the HETG provides the higher energy coverage of the 
warm absorber, especially in the 1 keV band and above and covering the Ne through Fe K-shell lines, with the RGS providing high sensitivity coverage at lower energies down to 0.3\,keV. 
 
The paper is organized as follows. In Section~2, we
  describe the observations and reduction of the HETG
  spectra. Section~3 is devoted to the spectral fitting of the HETG data, including 
atomic line detections and
  identifications as well as the measurements of the kinematics and profiles of the absorption lines. 
The properties of the emission lines, such as in the iron K-shell band and the He-like triplets, are also presented.
Section~4 presents
  photoionization modeling of the X-ray absorption in the HETG, which as will be shown 
requires multiple absorption components to cover the wide range of ionization states seen in the absorption 
spectrum. In
  Section~5, we discuss about the origins and
  infer some physical properties of the X-ray absorbing media observed in Mrk\,1040. 
The properties of the X-ray gas are then compared to the images and kinematics of the nuclear 
[O\,\textsc{iii}] emitting gas seen from archival {\it Hubble Space Telescope} ({\it HST}) observations.
As we will show, the X-ray absorber appears to show no evidence of outflow to high precision 
and the ionized gas may be in part associated with a failed wind located within $\sim 100$\,pc of the nucleus.

\section{Chandra Observations and Data Reduction}\label{sec:obs}


\chandra\ observed Mrk\,1040 four times between
24 September 2013 to 3 March 2014 (see Table 1 for a summary), with three of the 
four observations 
occurring within a one week period from 25 February 2014 and 3 March 2014. 
Spectra were extracted with the \textsc{ciao} package v4.7 \citep{fruscione2006}. 
Only the first order dispersed spectra were considered for both the MEG (Medium Energy Grating) and 
HEG (High Energy Grating) and the $\pm1$ orders for each grating were subsequently combined 
for each sequence. No significant spectral variability was observed between the four sequences, 
with only modest $\sim 5$\% variations in count rate (see Table 1).
Therefore the spectra were combined from all four sequences to yield a single 1st order spectrum 
for each of the MEG and HEG, yielding respective net source count rates of $0.457\pm0.002$\,s$^{-1}$ (from 0.8--5.0\,keV) 
and $0.238\pm0.001$\,s$^{-1}$ (from 1.2-9.0\,keV) 
respectively for a total exposure time of 197.8\,ks. 
The total counts obtained exceeded 90000 and 47000 counts
for MEG and HEG respectively, while the background count rate was negligible. 
The undispersed zeroth order image shows no evidence for extended X-ray emission in excess of 
the telescope point spread function.

The resultant time-averaged 
source spectra were subsequently binned at 3 different levels. These 
correspond to either at Half Width at Half Maximum (HWHM) of the spectral resolution
(i.e. $\Delta \lambda = 10$\,m\AA\ and 
$\Delta \lambda = 5$\,m\AA\ bins for 
MEG and HEG respectively), at the FWHM of the resolution, or at twice the FWHM resolution. Note that at 1\,keV (or 12.3984\AA), the HWHM binning corresponds to velocity intervals of
$\Delta v=240$\,km\,s$^{-1}$ and $\Delta v=120$\,km\,s$^{-1}$, per spectral bin, for the MEG and HEG respectively.
The former finer HWHM binning was adopted for the subsequent spectral fitting as well as for obtaining velocity profiles of the lines, while the 
latter more coarsely binned spectra (FWHM or twice FWHM) were used for an initial inspection of the data and 
for some of the spectral plots.
The MEG and HEG spectra were analyzed over the energy ranges of 0.8--5.0\,keV and 
1.2--9.0\,keV respectively. 
Below 0.8\,keV, the signal to noise of the MEG data was low due to the build up 
of contamination over time on the ACIS-S CCDs and was excluded from the subsequent fitting. 
A cross normalization constant between the MEG and HEG spectra has been included in all the 
subsequent fits, however in each case it is consistent with 1.0.
Note the overall fluxes, obtained against an absorbed power-law, correspond to $F_{0.5-2\, {\rm keV}}=1.1\times10^{-11}$\,erg\,cm$^{-2}$\,s$^{-1}$ and 
$F_{2-10\,{\rm keV}}=3.7\times10^{-11}$\,erg\,cm$^{-2}$\,s$^{-1}$ respectively.

The C-statistic \citep{cash1979} was 
employed in the subsequent spectral fits to the HETG, as although an overall large number of counts were obtained in the observations, towards 
the lower energy (longer wavelength) 
end of each grating spectrum the total source counts per bin drops below $N<20$ in some of the 
spectral bins, which is also true at the centroids of the deepest absorption lines in the 
spectrum.
In the case of $\chi^{2}$ minimization, this would lead to the continuum level being 
somewhat underestimated at soft X-ray energies.
All fit parameters are given in the 
rest frame of the quasar at $z=0.016652$ and are stated in energy units, adopting the 
conversion factor of 1\,keV being equivalent to 12.3984\AA\ in wavelength.
In all of the fits, a Galactic absorption of hydrogen column density of
$N_{\rm H}=6.7\times10^{20}$\,cm$^{-2}$ \citep{kalberla2005} was adopted, modeled with the 
``Tuebingen--Boulder'' absorption model ({\sc tbabs} in {\sc xspec}), see \citet{wilms2000}
and references therein for the cross--sections and abundances used in this model. 
Note that upon including molecular 
hydrogen, the Galactic column rises to $N_{\rm H}=8.84\times10^{20}$\,cm$^{-2}$, e.g. 
see \citet{willingale2013}, although the spectra above 0.8\,keV are not sensitive 
to the exact value adopted here.
Values of $H_{\rm 0}=70$\,km\,s$^{-1}$\,Mpc$^{-1}$,
and $\Omega_{\Lambda_{\rm 0}}=0.73$ are assumed throughout and errors are quoted at 90\% 
confidence ($\Delta\chi^{2}=2.7$), for 1 parameter of interest. 

\section{X-ray Spectral Fitting}\label{sec:initial}

\subsection{The Overall Spectral Form}
 
\begin{figure}
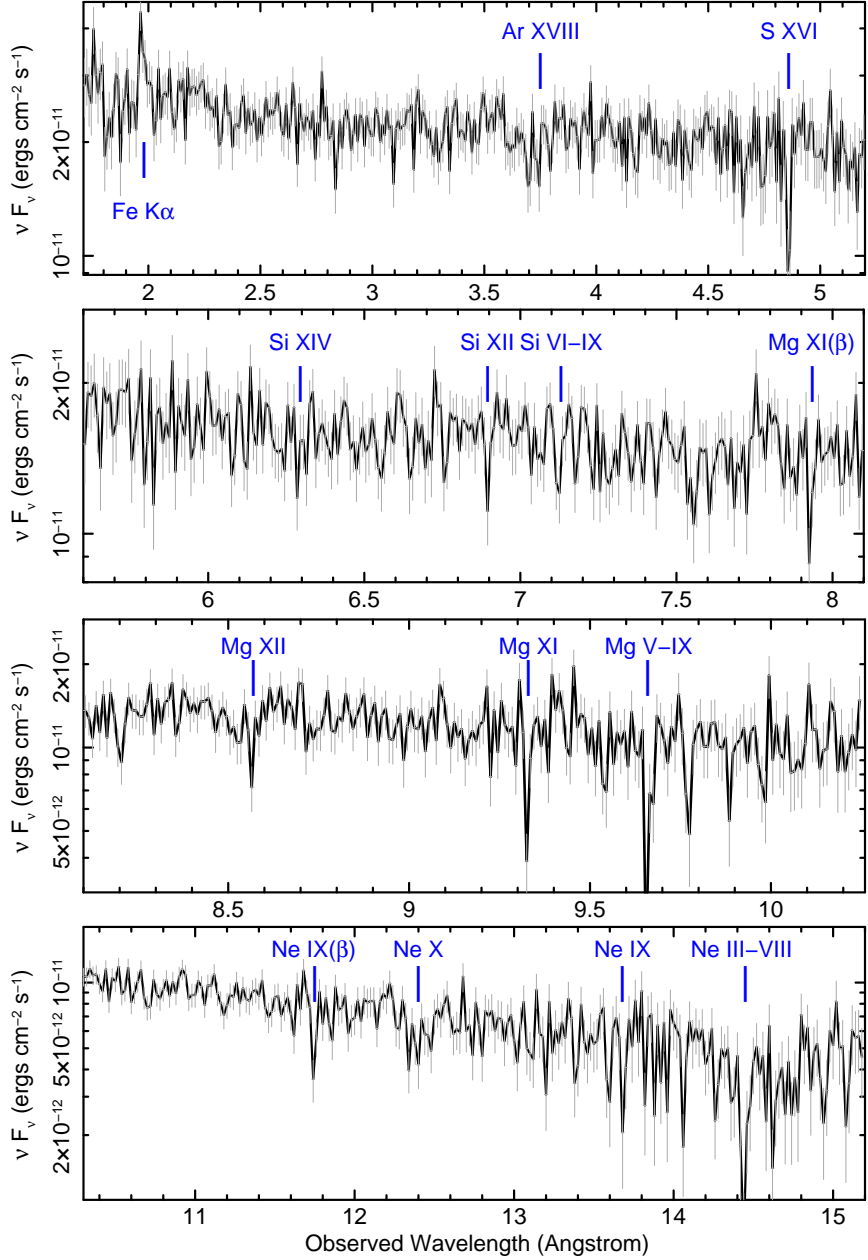

\begin{center}
\rotatebox{-90}{\includegraphics[height=12cm]{f1a.eps}}
\rotatebox{-90}{\includegraphics[height=12cm]{f1b.eps}}
\rotatebox{-90}{\includegraphics[height=12cm]{f1c.eps}}
\rotatebox{-90}{\includegraphics[height=12cm]{f1d.eps}}
\end{center}
\caption{The four panels show the {\it Chandra} HETG spectrum of Mrk 1040, 
in order of increasing wavelength, obtained from 
combining all 4 sequences in Table\,1. The upper three panels are from the HEG grating and the lower panel from the MEG. Error bars are shown in greyscale and the 
spectra are binned at FWHM resolution of $\Delta\lambda=10$\,m\AA\, and  $\Delta\lambda=20$\,m\AA\, for the HEG and MEG respectively. 
The spectra have been 
fluxed against a power-law continuum to create a $\nu F_{\nu}$ 
spectrum and are plotted in the observed frame. Multiple absorption lines from an ionized absorber are 
present in the spectrum, as seen from bottom to top:- (i) Ne\,\textsc{iii-x}, (ii) Mg\,\textsc{v-xii}, (iii) Si\,\textsc{vi-xiv} and (iv) S\,\textsc{xvi} and Ar\,\textsc{xviii}. Thus absorption is seen from ions 
covering a wide range of ionization, from the low ionization inner-shell lines of Ne, Mg and Si (see Behar \& Netzer 2002), through to the H-like lines.}
\label{fluxed}
\end{figure}

Figure\,\ref{fluxed} shows the observed HETG spectrum of Mrk\,1040, 
where the spectrum has been unfolded through the instrumental response 
against a powerlaw model of photon index $\Gamma=2$ 
in order to produce a fluxed spectrum in $\nu F_{\nu}$ flux units. The spectrum has been plotted 
at the FWHM spectral resolution and over the wavelength (energy) range from 1.7\AA--15.2\AA\ 
(or 0.82--7.3\,keV) and is plotted in wavelength (in the observed frame) to show the 
overall spectral form.
The spectrum shows many clear signatures of a warm absorber, with several 
K-shell absorption lines present from the most abundant elements, where the strongest lines 
have been identified in the figure panels. 
Aside from the highly ionized absorption lines from He and H-like ions 
(e.g. from Ne, Mg and Si, as well as the H-like lines from S and Ar), evidence for lower ionization 
absorption is also present. This is in the form of inner K-shell ($1s\rightarrow2p$) 
absorption lines from ions corresponding to 
charge states from Li to F-like (i.e. where the L-shell is partially occupied), see \cite{behar2002}.
These can be seen in the 
spectrum via absorption lines from charge states varying from:- Ne\,\textsc{iii-viii} (14.0--14.8\AA\ observed frame), 
Mg\,\textsc{v-ix} (9.5-10.0\AA) and from Si\,\textsc{vi-xii} (6.8--7.2\AA). The most prominent of 
these lines appear from Ne\,\textsc{v} (C-like Ne) and Mg\,\textsc{viii} (B-like Mg). This suggests that 
the warm absorbing gas covers a wide range of ionization, as has been observed in other 
Seyfert 1 galaxies, e.g. NGC\,3783: \citep{kaspi2000,kaspi2001,krongold2003}, NGC\,5548: \citep{kaastra2002,andrade2010}, Mrk\,509: \citep{yaqoob2003,smith2007}, NGC\,7469: \citep{blustin2007}, IC\,4329A: \citep{steenbrugge2005b}, NGC\,3516: \citep{holczer&behar2012},
MCG-6-30-15: \citep{lee2001,turner2004}, Mrk\,279: \citep{costantini2007}. 

Note that we do not detect absorption due to either the low ionization ions of iron, i.e. in the form of the 
iron M-shell Unresolved Transition Array (Fe\,\textsc{i-xvii} UTA, \citealt{sako2001,behar2001}), or due to Oxygen \citep{gu2005}, as these fall below 0.8\,keV where the 
effective area of the HETG drops rapidly. However the study of the absorption 
from these lower energies features will be presented in paper II covering the soft X-ray RGS 
spectrum.

\subsection{Absorption Lines in the HETG Spectrum}

In order to measure the absorption line properties, an initial parameterization of the continuum 
was adopted. The continuum emission was modified by a powerlaw absorbed by the Galactic column and a
neutral partial covering absorber was also included (using the \textsc{xspec} model \textsc{zpcfabs}), 
whereby a fraction $f_{\rm cov}$ of the primary X-ray continuum passes through the absorber.
This simple model is adopted
purely to account for the spectral curvature (decrease in flux) towards lower energies due to bound-free 
absorption, but without accounting for any discrete absorption lines. A full description 
of the photoionization 
modeling of the spectrum which accounts for both the absorption lines present and the spectral curvature 
is given in Section 4 and its effect on the overall spectrum is shown later in Figure 8.
The photon index of the powerlaw was found to be $\Gamma=1.75\pm0.02$, while the column of the partial 
covering absorber was found to be $N_{\rm H}=4.0^{+0.9}_{-0.3}\times10^{21}$\,cm$^{-2}$ with a line of sight covering fraction of $f_{\rm cov}>0.92$, i.e. consistent with a fully covering absorber.

\begin{figure}
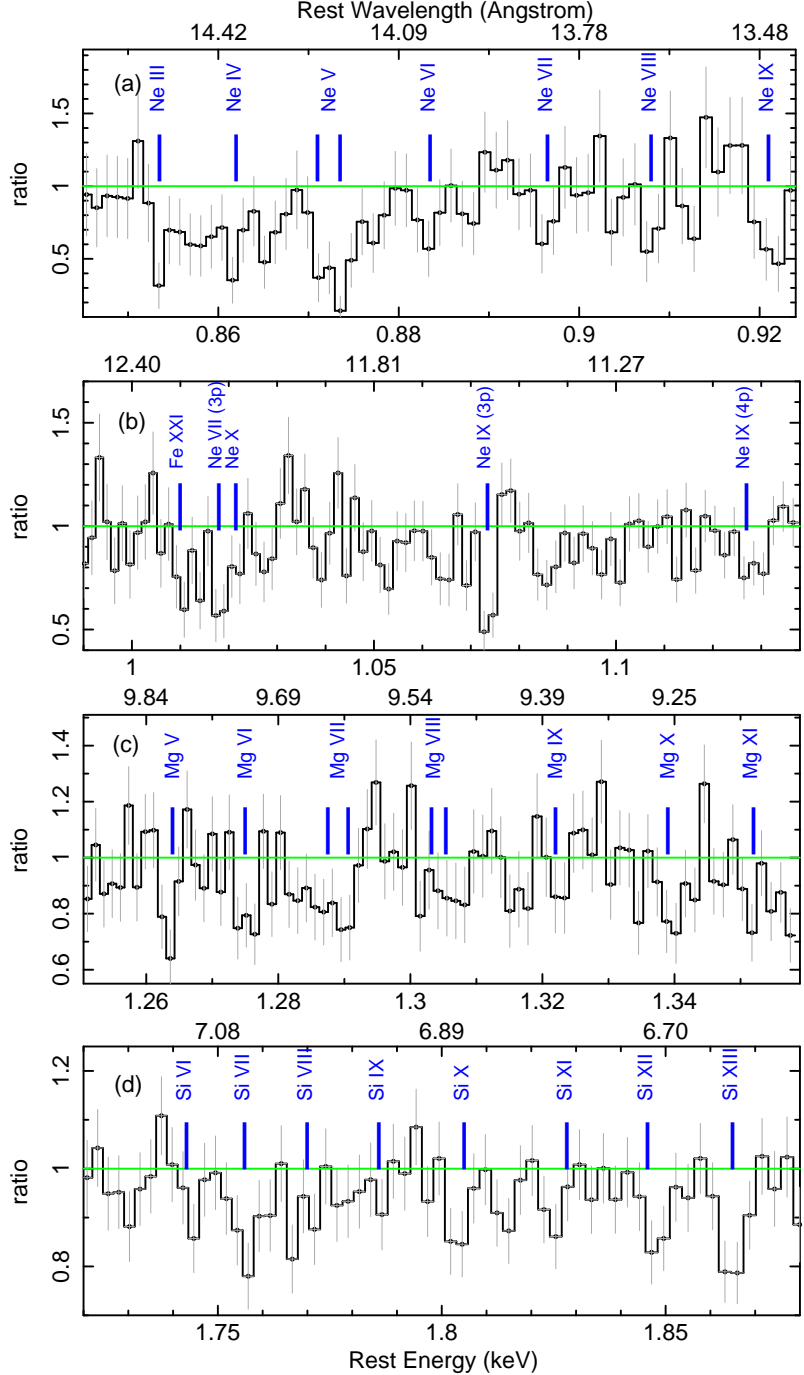

\begin{center}
\rotatebox{-90}{\includegraphics[height=11cm]{f2a.eps}}
\rotatebox{-90}{\includegraphics[height=11cm]{f2b.eps}}
\rotatebox{-90}{\includegraphics[height=11cm]{f2c.eps}}
\rotatebox{-90}{\includegraphics[height=11cm]{f2d.eps}}
\end{center}
\caption{A zoom-in showing the absorption line structure observed in Mrk\,1040. The four panels show the data/model residuals to the {\it Chandra} MEG spectrum, binned at 
FWHM resolution, with respect to the baseline continuum consisting of a powerlaw with neutral absorption, as described in Section\,3. 
The panels are plotted in the AGN rest frame (at $z=0.016652$), with energy shown on the lower x-axis 
and wavelength on the upper-axis. A wealth of absorption lines are seen in the spectrum, from top to bottom:- (a) Ne\,\textsc{iii-ix}, (b) Ne\,\textsc{ix-x}, (c) Mg\,\textsc{v-xi} and (d) Si\,\textsc{vi-xiii}. 
Thus as well as the highly ionized He and H-like lines, absorption is seen from a range of inner K-shell 
lines of low to intermediate ionization, varying from O or F-like ions to Li-like ions. See Table\,2 for 
details of the lines present in the spectra.}
\label{meg-panels}
\end{figure}

\begin{figure}
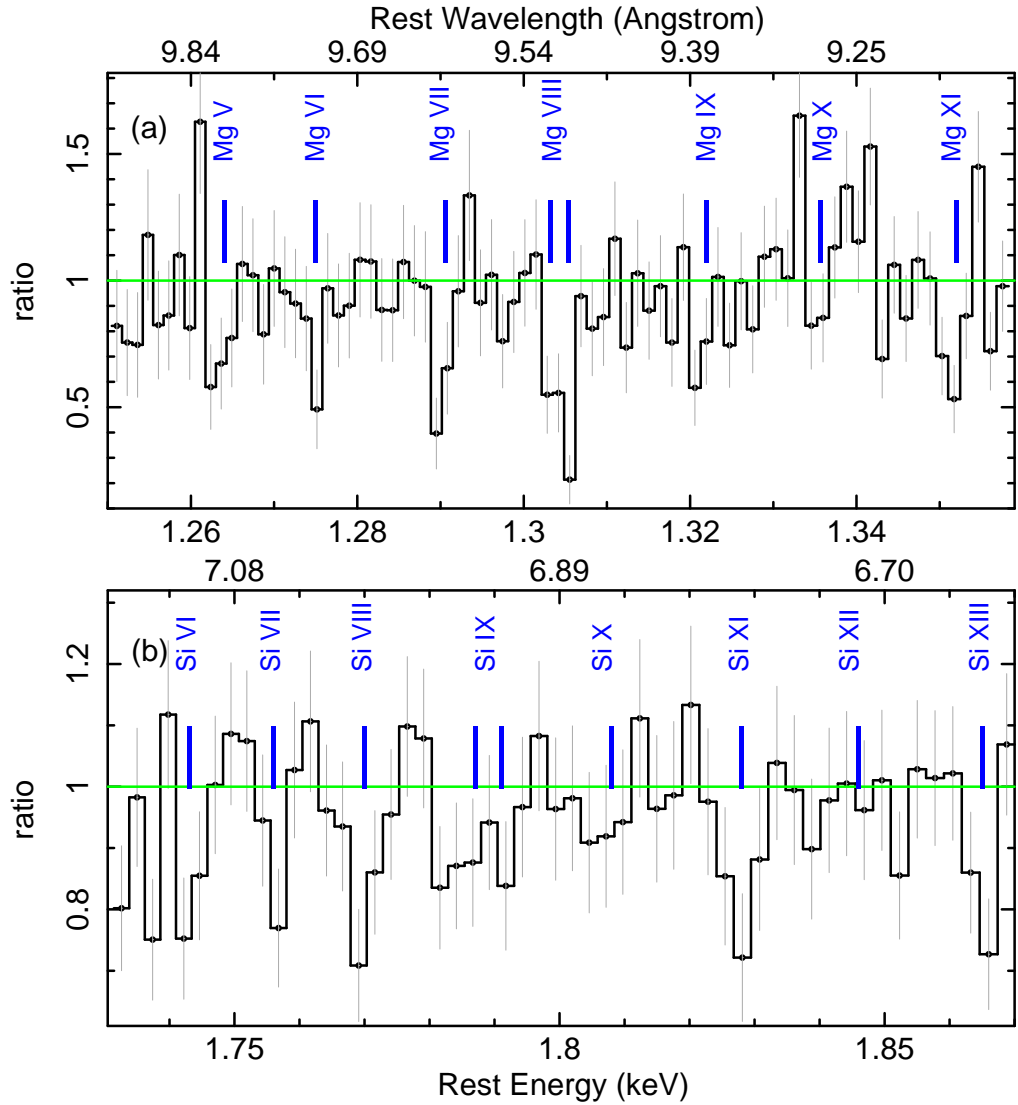

\begin{center}
\rotatebox{-90}{\includegraphics[height=14cm]{f3a.eps}}
\rotatebox{-90}{\includegraphics[height=14cm]{f3b.eps}}
\end{center}
\caption{As per Figure~2, but showing portions of the Chandra HEG spectrum of Mrk\,1040 at FWHM resolution. 
The upper panel shows the Mg K-shell series of lines, the lower panel the Si K-shell series, 
each for different 
ionic charge states. The vertical blue lines mark the expected rest energy (wavelength) of the absorption lines for each charge state from \cite{behar2002}.}
\label{heg-panels}
\end{figure}

While this simple partial covering model is successful in reproducing the spectral curvature, 
the overall fit is poor, with a fit statistic of $C/\nu=2592.7/2262$ (where $\nu$ is the number of 
degrees of freedom) which is rejected at $>99.99$\% confidence. The residuals to this 
continuum model in the Ne, Mg and Si bands to the MEG spectrum are shown in Figure\,\ref{meg-panels}, 
while the equivalent residuals in the Mg and Si bands to the HEG spectrum are shown in 
Figure\,\ref{heg-panels}. 
Indeed a wealth of absorption line structure is
clearly present in the HETG spectrum (independently in both the MEG and HEG gratings) against  
the continuum model, especially at energies below 2\,keV in the Ne, Mg and Si bands. 
As expected from the initial broad-band spectrum, 
low ionization gas appears to be present in the form of a multitude of inner K-shell lines 
of Ne, Mg and Si. We refer to
\cite{behar2002} for a compilation of 
these inner shell lines and we adopt the known energies (wavelengths) of these lines from this paper 
later in Table\,2. 
Indeed such lines have been detected in other high signal to noise
grating spectra of Seyfert 1 AGN, 
such as in NGC\,3783 \citep{kaspi2002,blustin2002},
  NGC 4151 \citep{kraemer2005}, Mrk 509 \citep{kaastra2011b}, NGC 3516
  \citep{holczer&behar2012}, NGC\,4051 \citep{lobban2011},
  NGC 5548 \citep{steenbrugge2005}, MR\,2251-178 \citep{reeves2013} and 
MCG\,-6-30-15 \citep{holczer2010} among others.

In the Ne band of the Mrk\,1040 MEG spectrum, residuals due to inner-shell 
absorption lines due to Ne\,\textsc{iii-vii} (O-like Ne through to Be-like Ne) are present from 
0.85-0.90\,keV (13.8\AA-14.6\AA). In addition at higher energies absorption from He and H-like 
Ne are also present, in particular the He-like line series shows absorption from $1s\rightarrow2p$ up to 
$1s\rightarrow4p$ (see Figure 2, panels a and b). This indicates the absorber exhibits a wide range of ionization contributing many different charge states to the absorption spectrum.
Likewise, the spectrum in the Mg and Si K-shell bands shows similar characteristics. 
Inner-shell lines from Mg\,\textsc{v-ix} (O-like through to Be-like ions) are apparent from $1.26-1.32$\,keV (9.4--9.8\AA). Similarly, 
inner shell absorption is also detected from Si\,\textsc{vi-xii} (F-like to Li-like) 
around 1.74--1.85\,keV (6.7--7.1\AA). Most of the lines are also independently
detected in the HEG (Figure\,\ref{heg-panels}) as  
well as the MEG (Figure\,\ref{meg-panels}) spectra. Absorption from the resonance ($1s\rightarrow2p$) 
He-like lines of Mg and Si are also detected (as well as the H-like absorption for Mg\,\textsc{xii}), 
which again confirms that ions from a wide range of charge states are present 
in the Mrk\,1040 absorber.

\subsection{Velocity Profiles}

We then constructed velocity profiles of the principle absorption lines listed in Table\,2.
We restricted the analysis to those lines which are detected at at least 95\% confidence 
(corresponding to $\Delta C=6.0$ for 2 interesting parameters), noting that most of the lines are detected 
at least at this level in both gratings (with the exception of the Ne band as noted below).
A baseline continuum consisting of an absorbed power-law was used, utilizing the 
neutral partial covering model described in the previous section, as this accounts for the 
broad-band spectral curvature without imparting any discrete absorption lines upon the spectrum. 
The lines were then fitted with Gaussian profiles in the rest energy frame, 
allowing their centroid energies, widths and normalisations to vary and accounting for the spectral 
resolution via the standard instrument responses.
To maximize the resolution, HWHM binning was used and we utilized the highest resolution 
HEG data for the Mg and Si bands (and above), except for the Ne band where the MEG is adopted due to the low effective area of the HEG at 1\,keV and below.
The line profiles were then transposed into velocity space 
around the known lab frame energy (or wavelength) of each line.
In the case where a line profile originates from an unresolved doublet (such as for the H-like lines), we took the weighted mean of the respective lab frame energies to calculate the expected line energy. 
The doublet corresponding to the 
Mg\,\textsc{viii} $1s\rightarrow2p$ lines is resolved by the HEG, so in that case we fitted the two lines separately, 
obtaining the centroid in velocity space separately for each component, i.e. allowing for the 
different rest-energies of the two lines.

The results of the line profile fitting are summarized in Table\,2, which includes 21 line profiles 
covering 18 separate ionic species from Ne up to Ar. 
The absorption profiles arise mainly from the $1s\rightarrow2p$ transitions, 
except for Ne\,\textsc{ix} and Ne\,\textsc{x}, where the higher order $1s\rightarrow3p$ lines are detected; indeed 
in these latter cases the He-$\beta$ and Ly-$\beta$ profiles provide better constraints on the line 
widths and velocity centroids, as they are free from contamination from other lines.
It is also apparent from inspecting the measured energy centroids in Table\,2 that most of the lines are positioned close to their expected lab frame energies, while generally the absorption 
lines are unresolved (or only marginally resolved), with typical velocity widths of $\sigma<300$\,km\,s$^{-1}$ (or even $\sigma<100$\,km\,s$^{-1}$ in the more tightly constrained cases).  

Examples of some of the velocity profiles are plotted in Figure\,4 
(for the inner-shell and He-like resonance lines) and in Figure\,5 (for the H-like lines). 
In all of the examples, the line centroids are consistent with zero velocity shift within the measurement errors. While it may appear that some of the profiles contain an additional higher velocity component (either red or blue-shifted) with respect to the main line centroids, these are due to the presence of nearby lines from adjacent ions as labelled in 
Figures 4 and 5. For example the 
strong Ne\,\textsc{v} absorption profile shows weak absorption components at $\pm4000$\,km\,s$^{-1}$; these are simply due to neighbouring absorption lines from Ne\,\textsc{iv} and 
Ne\,\textsc{vi} either side of Ne\,\textsc{v} with no velocity shift.
This is also the case for the Si\,\textsc{viii} and Si\,\textsc{xiii} 
profiles, e.g. the latter He-like profile contains a weak contribution from Si\,\textsc{xii} (Li-like Si) on the red-side of the profile. The velocity widths of these profiles are generally low or unresolved, only Ne\,\textsc{v} is significantly broadened (with $\sigma=490\pm200$\,km\,s$^{-1}$). However this may be a result of the line being an unresolved doublet \citep{behar2002}, or from this deep line becoming saturated in its core (as the flux reaches close to zero at the center). 
In contrast the He-like profiles of Ne\,\textsc{ix} (He-$\beta$) and 
Mg\,\textsc{xi} (neither of which are doublets, saturated, or contaminated by other lines) are only marginally 
resolved, with widths of $\sigma=225^{+140}_{-100}$\,km\,s$^{-1}$ and 
$\sigma=150^{+110}_{-95}$\,km\,s$^{-1}$, respectively. The Mg\,\textsc{viii} profile is clearly resolved 
in the HEG data into a doublet, although the velocity shifts and widths of both components are consistent with zero (e.g. $\sigma<100$\,km\,s$^{-1}$).

A similar story emerges for the H-like (Lyman-$\alpha$) profiles, as shown in Figure 5 for Ne\,\textsc{x}, 
Mg\,\textsc{xi}, S\,\textsc{xvi} and Ar\,\textsc{xviii} (note Si\,\textsc{xiv} falls just below the detection 
threshold). The Ne\,\textsc{x} Ly-$\alpha$ profile is heavily contaminated by $2p\rightarrow3d$ absorption 
by Fe\,\textsc{xxi} and thus the velocity constraints listed in Table\,2 are instead 
obtained from Ne\,\textsc{x} Ly-$\beta$. 
The highly ionized S\,\textsc{xvi} and Ar\,\textsc{xviii} profiles are both unresolved and their velocity 
centroids are consistent with zero. Only the Mg\,\textsc{xii} profile contains some indication of 
blueshift. Although the main centroid is consistent with zero velocity, a second velocity component 
emerges at $-1500$\,km\,s$^{-1}$ which appears uncontaminated by other lines (note that the 
separation of the Mg\,\textsc{xii} doublet cannot account for this as it is too small to be resolved by the HEG). However upon inspecting the 
MEG data for consistency, no such blueshifted component was confirmed, nor is such a component detected in any of the other high ionization He or H-like lines. Even though this component appears formally significant at the $\sim99$\% level ($\Delta C=10.6$), as no high velocity component was measured in 
any of the other profiles we do not consider this detection to be plausible.

Overall none of the individual profiles, low ionization as well as high ionization (He and H-like), appear to show a significant outflowing component and most of the lines are unresolved within the instrumental resolution. To determine if the absorption profiles as a whole are characterized by a small velocity shift, which 
is within the statistical error of any individual profile, the mean and dispersion of the velocity shifts of the 
sample of lines was calculated based on the values reported in Table\,2. Only one value was considered for 
each ionic species, taking the best determined values in the three cases where multiple measurements exist
for a given ion (e.g. the Ne\,\textsc{x} $1s\rightarrow3p$ line was favored 
over the $1s\rightarrow2p$ line). Over 18 ions, 
the mean velocity shift was found to be $<v_{\rm out}>=0\pm40$\,km\,s$^{-1}$, with a 
dispersion of $\sigma_{v}=180$\,km\,s$^{-1}$. Neither was any trend or deviation 
found between outflow velocity and rest energy. Thus no outflow (or inflow) is required to within 
$\pm40$\,km\,s$^{-1}$ of the systemic velocity of Mrk\,1040, within the absolute 
wavelength calibration of the HETG gratings. Thus unlike for many other AGN, the Mrk\,1040 warm absorber does not appear to originate from an outflow, at least to within very tight limits. 
In the next section we return to 
model the absorption with photoionization models generated by \textsc{xstar}.

\subsection{Emission Lines}

In contrast to the absorption lines, there are a paucity of soft X-ray emission lines in the HETG spectrum 
against the bright X-ray continuum. The emission lines which are detected in the spectrum are 
listed in Table\,3.
In the soft X-ray band the most significant emission residuals occur 
around the Mg\,\textsc{xi} triplet, at the expected positions of the forbidden and intercombination 
transitions, while an absorption line is present at the resonance line energy as is shown in Figure\,6 
for the HEG data. Both emission lines are relatively weak (equivalent width of $1.1\pm0.7$\,eV) 
and appear unresolved, with an upper limit to their velocity widths of $\sigma<135$\,km\,s$^{-1}$. 

The relatively equal contribution of the forbidden and intercombination components 
appears surprising at first, as this would imply a relatively low ratio of the $R$ value between the two lines ($R=z/(x+y)$, see \citealt{porquet2000}) and subsequently a 
high density of $n_{\rm e}\sim10^{13}$\,cm$^{-3}$. However given the large errors on the line 
normalizations it is not possible to constrain the line ratio and the derived density is also 
formally consistent with 
a lower density photoionized plasma, where $n_{\rm e}<10^{12}$\,cm$^{-3}$. 
Furthermore a relatively high UV radiation field may also suppress the forbidden line, 
via photoexcitation \citep{porquet2010}, making the direct measurement of the density 
from the triplet lines less certain. Furthermore the forbidden line
may also be absorbed on its blue-wing due to the adjacent Mg\,\textsc{x} 
resonance absorption line (see Figure 6). 
Note that no other line emission is detected from the other triplets, e.g. as the S/N is too low around 
Ne\,\textsc{ix}, while 
the only other soft X-ray line which is marginally detected arises from Si\,\textsc{xiv} Lyman-$\alpha$ 
at 2.0\,keV, which again is unresolved.

\subsubsection{The Iron K-shell band}

The strongest signal present in the iron K-shell band arises from the almost ubiquitous narrow 
Fe K$\alpha$ emission line \citep{nandra2007}; the iron K band spectrum is plotted in Figure\,7.
The line is detected at $6403^{+12}_{-8}$\,eV and thus is at the expected energy for near neutral 
iron. The equivalent width of the narrow Fe K$\alpha$ line is $44^{+17}_{-12}$\,eV. 
The line is unresolved, with a formal upper-limit to its width of $\sigma<28$\,eV or 
$\sigma<1310$\,km\,s$^{-1}$ in velocity space. This appears consistent with an origin in distant 
(e.g. pc scale) gas, which is often found to be case from Fe K$\alpha$ line cores seen 
in other type I Seyferts as measured by Chandra HETG \citep{shu2010}. Note there is some indication of an excess red-wards of the line core, 
this is consistent with the presence of a Compton shoulder to the line \citep{george&fabian1991} 
with an energy of $E=6.33^{+0.03}_{-0.07}$\,keV; 
however formally the Compton shoulder is not detected ($\Delta C=4.3$) and 
neither is it well constrained. Aside from the possibility of a weak Compton shoulder, there is 
no evidence for any additional excess emission on the red-wing of the line, which may be expected to arise 
from a broad line from the inner accretion disk. 
The presence (or lack thereof) of any relativistic line component 
in Mrk\,1040 will be discussed in a forthcoming paper.

Aside from the neutral K$\alpha$ emission, there is some evidence (at $\Delta C=6.3$) for the 
presence of ionized emission exactly coincident with the expected position of the 
forbidden line from He-like Fe (Fe\,\textsc{xxv}). Although weak ($EW=17\pm11$\,eV), 
the line centroid is accurately measured to $E=6630^{+15}_{-6}$\,eV, consistent with the known 
energy of the forbidden component and the line width appears narrow with $\sigma<14$\,eV 
(or equivalently $\sigma<630$\,km\,s$^{-1}$). Thus its origin likely arises from distant photoionized 
material, as per the soft X-ray line emitting and absorbing gas.
Note that the residuals at the expected position of the resonance absorption lines from 
Fe\,\textsc{xxv} and Fe\,\textsc{xxvi} (6.70\,keV and 6.97\,keV) 
are not formally significant in their own right ($\Delta C=4.4$ and 
$\Delta C=3.7$ respectively), however in the next section we will place a formal limit on any high 
ionization absorption component arising from Mrk\,1040.

Finally in order to determine whether the observed narrow iron K$\alpha$ emission is consistent 
with an origin in scattering off distant Compton thick matter, we instead modeled with spectrum 
with either a neutral or an ionized Compton reflection component in addition to the underlying 
power-law continuum. An excellent fit ($C/\nu=565.2/512$ over the 2--8\,keV band) 
was found from modeling the line with a 
neutral reflection component, adopting the \textsc{pexmon} model \citep{nandra2007} 
within \textsc{xspec}, which returned a reflection fraction of $R=0.30^{+0.06}_{-0.12}$ for 
an assumed inclination angle of $\theta=45$\,degrees and Solar abundances. 
Note that the photon index of the 
underlying continuum was $\Gamma=1.79^{+0.07}_{-0.04}$. A statistically equivalent 
fit was also obtained with an ionized reflection model, using the (angle averaged) 
tabulated \textsc{xillver} 
model spectra of \cite{garcia2010} and \cite{garcia2013}, where the upper-limit on the ionization parameter 
was found to be $\log \xi<0.1$, with a consistent photon index of $\Gamma=1.88^{+0.05}_{-0.08}$. Indeed the spectrum fitted with this model is shown in Figure\,7.
Thus the Fe K$\alpha$ emission appears consistent with an origin from distant, Compton thick 
matter and no additional velocity broadening is required to account for this band.

\section{Photoionization Modeling of the X-ray Absorption Spectrum}\label{sec:absorption}

\subsection{Photoionization Models}

Given the substantial presence of partially ionized gas in the X-ray spectrum of Mrk\,1040, 
we attempted to model the 
absorption spectrum with photoionized grids of models using the \textsc{xstar} code v2.2 
\citep{kallman2004}. 
Absorption grids were generated in the form of \textsc{xspec} multiplicative tables. 
The absorption spectra within each grid were computed between 0.1--20\,keV with $N=10000$ spectral bins. 
The photoionizing X-ray continuum between 1--1000\,Rydberg was assumed to be a broken 
power-law, with a photon index of $\Gamma=2.5$ below a break energy of 0.8\,keV (at the low energy end 
of the HETG bandpass) and with a 
photon index of $\Gamma=2$ assumed above this break energy. 
The 1--1000\,Rydberg photoionizing luminosity is predicted to be 
$L_{\rm ion}\sim2\times10^{44}$\,erg\,s$^{-1}$ for Mrk\,1040 using this continuum model.
Although the power-law break is not required to model the continuum in the HETG data, which is 
not sensitive to the lowest energies,
it is required by the subsequent \xmm\ spectra (both for the pn and RGS, paper II) to account 
for the soft X-ray excess in this source. Furthermore, as will be discussed in 
Section\,4.3, a softer photoionizing continuum is required to model the 
numerous low ionization inner-shell absorption lines in the spectrum, which 
otherwise are over ionized by a flatter $\Gamma=2$ continuum. 

With this choice of continuum, we then generated an initial broad grid of models that covered a wide range in ionization and column density:- 
from $N_{\rm H}=1\times10^{18}$\,cm$^{-2}$ to $N_{\rm H}=3\times10^{24}$\,cm$^{-2}$ 
and $\logxi = 0-5$\footnote{The ionization parameter is defined as $\xi=L_{\rm ion}/nR^{2}$ (\citealt{tarter1969}), where $L_{\rm ion}$ is the
$1-1000$ Rydberg ionizing luminosity, $n$ is the electron density
and $R$ is the distance of the ionizing source from the absorbing
clouds. The units of $\xi$ are erg\,cm\,s$^{-1}$.} in logarithmic steps of $\Delta (\log N_{\rm H}) = 0.5$ and 
$\Delta (\log \xi) = 0.5$ respectively. 
This broad grid was used to provide an initial characterization of the warm absorber zones (i.e. to provide a first order estimate of the range of ionization and column needed to account for the absorption) 
and to account for any high ionization absorption components. A more finely tuned grid (covering 
a narrower range of parameters) was subsequently generated with the 
specific purpose of modeling the low ionization absorption in the Mrk\,1040 spectrum, especially the inner-shell lines. The column density of this narrow low ionization grid 
covered the range from $N_{\rm H}=0.5-5.0\times10^{21}$\,cm$^{-2}$ in steps of 
$\Delta N_{\rm H} = 1\times10^{20}$\,cm$^{-2}$, with the ionization range extending from 
$\logxi = -1-3$ in 20 steps of $\Delta (\log \xi)=0.2$. 
A fine spectral resolution of $N=10^{5}$ points over an energy range of $0.1-20$\,keV was 
also employed. 
Given the narrow (or unresolved) widths of the absorption lines detected in the 
\chandra\ HETG spectrum, a turbulence velocity\footnote{The turbulence velocity width is defined as $b=\sqrt{2} \sigma = {\rm FWHM}/(2\sqrt{\ln 2})$.} 
of  $b=100$\,km\,s$^{-1}$ was assumed for the 
absorption models; grids with much higher turbulences all gave substantially
worse fits in the models considered below. 
Solar abundances were adopted for all the abundant elements, using the
values of \cite{grevesse1998}, 
except for Ni which is set to zero (the default option within \textsc{xstar}).

\subsection{Warm Absorber fits to the Mrk\,1040 \chandra\ HETG spectrum}

In order to model the absorption spectrum we successively added individual 
components of absorbing gas, fully covering the line of sight to the source, until the fit statistic was  
no longer improved at the 99\% confidence level ($\Delta C=9.2$ for 2 parameters)
and no obvious residuals remained after fitting the spectrum.
To fit the continuum itself we adopted the broken powerlaw form as described above, with the break 
energy fixed at 0.8\,keV and the photon index below the break also fixed at $\Gamma=2.5$ 
(as the Chandra data are not sensitive to the slope of the soft excess). 
Note that the above choice of the continuum is consistent with the later \xmm\ observation, 
where the low energy soft excess is required.
Above the break energy the 
photon index was allowed to vary, along with the overall normalization of the continuum.
After adding the required absorption zones as described below, the photon index 
was found to be $\Gamma=1.78\pm0.02$, while after correcting for the absorption, 
the intrinsic luminosity of Mrk\,1040 was $L_{2-10}=2.4\times10^{43}$\,erg\,s$^{-1}$ and 
$L_{\rm 1-1000\,Ryd}=1.8\times10^{44}$\,erg\,s$^{-1}$ over the 2--10\,keV and 1--1000\,Rydberg 
bands respectively.
The emission lines were also added to the model as Gaussians, as described in Section\,3.4; these 
arise from the Mg\,\textsc{xi} triplet, from the Fe K$\alpha$ and Fe\,\textsc{xxv} components of the 
iron K band emission and a weak emission line from Si\,\textsc{xiv} Lyman-$\alpha$ (see Table\, 3). 

Three components of absorbing gas are formally required  
in the \chandra\ model to account for the soft X-ray absorption, which are listed as zones\,1--3 in Table\,4. 
The finely tuned grid of absorption models was used for all three soft X-ray zones.
A fourth highly ionized absorption zone was also added to the model (zone 4, Table\,4), 
using the coarse grid covering a broader 
range of column and ionization. However compared to the soft X-ray zones 1--3, this high ionization zone 
is only marginally significant ($\Delta C=11.0$) and mainly contributes by adding the weak absorption lines due 
to Fe\,\textsc{xxv} and Fe\,\textsc{xxvi} (see Section 3.4.1 and Figure 7).
Overall with all the zones of absorption applied to the continuum, the fit to the \chandra\ spectrum is 
very good, with a final fit statistic of $C/\nu = 2374.0/2252$ and a corresponding null hypothesis probability of 
$P_{\rm f}=0.16$. 
In comparison the fit with the simple neutral partial covering model yielded a
substantially worse fit with $C/\nu = 2592.7/2262$ (as this model only accounts for the continuum curvature but 
not the line absorption), while a fit without any absorption zones applied to the continuum (aside from the 
Galactic column) is extremely poor with $C/\nu = 3508.3/2264$. Indeed the effect of the warm absorber upon 
the overall spectrum is illustrated in Figure 8. Here the top panel shows the spectrum fitted with just a 
power-law continuum (with $\Gamma=1.78$) above 3 keV, where the pronounced spectral curvature due to the 
warm absorber is apparent towards lower energies. Once the warm absorber is applied to the spectral model, 
then the continuum curvature is clearly accounted for by the absorbing gas (see Figure 8b).

Overall the three soft X-ray warm absorber zones\,1--3 that are required to model the 
\chandra\ spectrum cover the range in column from $N_{\rm H}=1.5-4.0\times10^{21}$\,cm$^{-2}$ 
and ionization from $\log\xi = 0-2$ (see Table\,4 for detailed parameters), where the 
lowest ionization zone 1 ($\log\xi=0$) has the highest column. 
As is expected from the velocity profile analysis (Section 3.3), the outflow velocities of the absorption 
zones are very low and are typically consistent within $\pm100$\,km\,s$^{-1}$ of the systemic velocity of 
Mrk\,1040. While the lowest ionization zone\,1 hints at some mild outflow (with $v_{\rm out} = -150^{+105}_{-100}$\,km\,s$^{-1}$), this is not supported by either zones 2 or 3; e.g. zone 2 
is consistent with zero ($v_{\rm out} = +10^{+80}_{-90}$\,km\,s$^{-1}$) and zone 3 suggests 
very mild inflow (with $v_{\rm out} = +130^{+70}_{-60}$\,km\,s$^{-1}$). 
However any slight difference 
in velocities between the zones are likely to be within the scatter of the measurements. 
Indeed even the highly ionized zone\,4 (where $\log\xi=3.7$) shows no evidence for outflow, 
with a formal limit on the outflow velocity of $<80$\,km\,s$^{-1}$.

Figure\,9 shows the relative contributions of each of the warm absorber components against a power-law 
continuum. The two lower ionization zones 1 and 2, with $\log\xi=0$ and $\log\xi=1$ respectively (top and middle panels), contribute to the lower ionization 
ions, reproducing the inner-shell lines of Ne, Mg and Si as 
seen in the spectrum below 2\,keV. Most of the bound-free 
spectral curvature is also imparted upon the spectrum by these two zones, with the lowest ionization 
zone\,1 having the greater opacity towards lower energies. 
The higher ionization zone\,3 (with $\log\xi=2.1$) mainly produces the He and H-like absorption lines
from Ne, Mg and Si, while the very highly ionized zone\,4 is essentially transparent at soft X-rays and only 
makes a weak contribution to some of the H-like lines. The superposition of these zones is then able to 
model the absorption line structure seen in the \chandra\ spectrum, where Figure\,10 shows the 
best-fit model over the 0.8-2\,keV band covering the major absorption lines from Ne, Mg and Si. 
Indeed the \textsc{xstar} model is able to account for the wide range of charge states 
present, e.g. reproducing 
the series of absorption lines observed from Ne\,\textsc{v-x}, Mg\,\textsc{vi-xii} and Si\,\textsc{vii-xiii}, while 
simultaneously being able to model for the convex shape of the overall X-ray spectrum.

The broad, and relatively flat distribution of $N_{\rm H}$ with $\xi$ seen in Table~4 is commensurate with other slow Seyfert outflows; see \cite{Behar2009} 
where this was parametrized as $N_{\rm H} \propto \xi ^a$.
Here the value of $a$ corresponds to a density profile of $n \propto r^{- \alpha}$, 
where $\alpha = (1+2a)/(1+a)$. 
Formally, the values in Table~4 (for zones 1--3) 
yield $a=-0.2\pm0.1$ and hence $\alpha=0.8\pm0.2$.
Such a flat distribution can be explained by an MHD outflow with $n \propto r^{-1} $ 
\citep{Fukumura2010}, but also with a Radiation-Pressure Compressed (RPC) cloud 
\citep{Stern2014}. 
The MHD wind models predict a well-defined line-of-sight outflow velocity 
structure of $v \propto r^{-1/2}$, that for a flat ionization 
distribution implies $v \propto \xi ^{1/2}$ that is not quite observed here.
The RPC models do not solve for the kinematic structure of the outflow and assume the entire absorber is cruising uniformly at the same velocity (that does not need to be specified for the ionization distribution to be flat). 
The latter appears to be the case for the absorber of Mrk 1040.
Clearly, the broad distribution of $N_{\rm H}$ with $\xi$ is not consistent with a simple radial outflow of $n \propto r^{-2}$, which would naturally produce a narrow distribution of ionization.

\subsection{The Effect of the Ionizing SED on the Absorber}

As discussed above, the shape of the illuminating SED from 1--1000\,Ryd may effect the 
properties of the X-ray absorber via the ionization balance. 
Here we adopted a broken powerlaw for the SED, with a steep spectrum below $\sim 1$\,keV, 
motivated by the need to model the lower ionization lines of Ne, Mg and Si in particular. These inner shell ions have relatively low ionization 
potentials for their L-shell electrons, e.g. 266\,eV for Mg\,\textsc{viii} and 126\,eV 
for Ne\,\textsc{v}; both of the ions produce notable absorption in the spectra as 
is shown in Figures 2 and 3. The low ionization potentials of these ions 
suggest that the form of the EUV and soft X-ray continuum may have an effect on the 
warm absorber modeling. 

In comparison we also tested the effect of a flat $\Gamma=2$ continuum on the absorber properties, 
by replacing the above grids of models with their equivalent versions with the flat SED shape 
and subsequently minimizing the new spectral fit.
In this case the 1--1000 Rydberg ionizing luminosity is a factor of 3 lower than for the 
broken power-law case ($5.6\times10^{43}$\,erg\,s$^{-1}$ vs. $1.8\times10^{44}$\,erg\,s$^{-1}$). 
The effect of the flat SED on the subsequent warm absorber modeling is quite apparent, 
which is illustrated in Figure\,11 in the Ne and Mg K-shell bands. The flat SED 
model clearly underpredicts the inner shell absorption from Mg\,\textsc{v-ix} and 
Ne\,\textsc{v-viii}, presumably due to the lower luminosity in the EUV to soft X-ray range and 
as an excess of such photons above the X-ray powerlaw is 
needed to excite these L-shell electrons. On the other hand, both models are able to account 
for the higher ionization absorption lines from He and H-like ions, as these are more sensitive
to the hard powerlaw part of the continuum above 1\,keV than the soft excess. 
Overall the fit with the flat SED model is significantly worse than the softer broken powerlaw 
SED model, with $C/\nu = 2452.6/2252$ vs. $C/\nu = 2374.0/2252$ respectively; 
here the difference in fit 
statistic arises from the inability of the former model to account for the lower ionization 
absorption lines. This demonstrates that the low ionization warm absorber is particularly 
sensitive to the form of the ionizing continuum in the lowest energy band; this has been found to be  
the case for the QSOs, MR\,2251-178 \citep{reeves2013} and IRAS\,13349+2438 \citep{laha2013}. 

Thus potentially the properties of the warm absorber may make it possible to deduce the 
nature of the soft X-ray continuum and especially the soft excess 
in AGN. While this is unlikely to be the result of the direct thermal emission from the accretion disk 
\citep{gierlinski2004}, some authors have suggested that the origin 
of the soft excess was atomic and may instead arise from an ionized reflector. 
An ionized reflection spectrum \citep{ross2005}, 
when convolved with the relativistic 
blurring expected in the innermost disk a few gravitational radii from the black hole, 
can produce an overall smooth continuum 
capabale of fitting the featureless soft excess below 1\,keV \citep{crummy2006,walton2013}. 
Alternatively, Comptonization of UV disk photons in a warm, optically thick, 
accretion disk atmosphere \citep{done2012,rozanska2015} has 
also proven to be a plausible form for the EUV to soft X-ray excess in 
many AGN \citep{jin2012,petrucci2013,matt2014}. 
In Mrk\,1040, the warm absorber modeling favors a steep ($\Gamma>2$) 
continuum from the EUV to soft X-rays. This might imply that the soft excess 
is part of the intrinsic continuum in this AGN and may be produced through disk Comptonization, 
rather than arising from 
reprocessed or reflected emission, as is also likely to be the case in MR\,2251-178 \citep{nardini2014}. 

\section{Discussion}\label{sec:discussion}

\subsection{Main observational results}

We have presented the first high resolution X-ray spectrum of the nearby, X-ray bright, Seyfert 1 galaxy, Markarian\,1040. 
The \chandra\ HETG observations reveal a spectrum showing a wealth of spectral features from the warm absorber, 
which covers a wide range in ionization; e.g. from lowly ionized O-like ions up to highly ionized He/H-like like species. The warm absorber parameters 
for this AGN, with column density in the range from $(1.5 < N_{\rm H} < 4.0)\times10^{21}$\,cm$^{-2}$ and 
$0 < \log\xi <2.1$, are fairly typical for most nearby Seyfert galaxies \citep{crenshaw2003,blustin2005,mckernan2007}, 
although at the lower range in ionization. 
Perhaps the most surprising discovery is the low (or zero) velocity of the gas, consistent within $\pm100$\,km\,s$^{-1}$ of the systemic 
velocity of Mrk\,1040. Thus unlike the majority of Seyfert 1 galaxies, whereby the warm absorbing X-ray gas is clearly ascribed to an outflow, 
the warm absorber in Mrk\,1040 either does not appear to be associated with an outflow, or instead 
any wind is not able to accelerate out to sufficient distances from the nucleus.

Indeed the velocities of most X-ray warm absorber components typically fall in the range from 0 up to $-2000$\,km\,s$^{-1}$ \citep{blustin2005,mckernan2007,laha2014}, while 
the UV velocity components can also cover a similar range of velocity as seen in X-rays \citep{crenshaw&kraemer2012}. 
Typical outflow velocities of a few hundred km\,s$^{-1}$ are also seen for the soft X-ray absorber in many well studied cases; e.g. NGC 3783 \citep{kaspi2002}, Mrk 509 \citep{kaastra2011b,detmers2011}. 
Furthermore, Seyfert 1 AGN often show multiple velocity components in the X-ray band, as is also commonly observed in the UV. 
For example in an analysis of a deep observation of MCG\,--6-30-15 observed with \chandra\ HETG, \cite{holczer2010} 
detect a low velocity absorber with $v_{\rm out}=-100\pm50$\,km\,s$^{-1}$, while a higher velocity 
absorber emerges with $v_{\rm out}=-1900\pm150$\,km\,s$^{-1}$ only in the highest ionization component of the warm absorber. In this case, this was directly seen in  
some of the H-like lines (e.g. Mg\,\textsc{xii} and Si\,\textsc{xiv}), where the high velocity component was revealed in the blue-wing of the velocity profiles for these 
lines. While the lower velocity component of the warm absorber in MCG\,--6-30-15 appears 
very similar to what is observed here (with a wealth of inner K-shell and UTA absorption present), no high velocity component of the 
high ionization lines is seen in Mrk\,1040 (e.g. see the velocity profiles in Figures 4 and 5). 

The higher ionization zone\,3 warm absorber in Mrk\,1040 (see Table 4) shows no evidence of outflow ($v_{\rm out}=+130^{+70}_{-60}$\,km\,s$^{-1}$), while the 
highest ionization zone 4 (mainly arising from the weak contribution of He and H-like iron) has an upper limit of $<80$\,km\,s$^{-1}$ to the outflow velocity. This 
is in contrast to the warm absorbers seen in many other Seyferts, where higher velocity zones can emerge in the higher ionization gas. 
For instance in NGC\,3516, fast outflowing zones of up to $-4000$\,km\,s$^{-1}$ were present when the AGN continuum brightened 
\citep{turner2008,holczer&behar2012}, which in this case may have coincided with the emergence (or illumination) of an inner disk wind component to the absorber. 
Similarly in NGC\,4051, both high (several thousand km\,s$^{-1}$) and low velocity (several hundred km\,s$^{-1}$) 
warm absorbing zones appear to be present \citep{steenbrugge2009,lobban2011,pounds&vaughan2011}, while in NGC\,5548 the onset of a 
high velocity (BAL-like) component of the UV and X-ray absorber was also recently revealed in during an absorption episode \citep{kaastra2014}. 

Thus Mrk\,1040 appears curious in that there is no evidence for the X-ray warm absorber to be outflowing, either through a slow component (with $v_{\rm out}>100$\,km\,s$^{-1}$) 
or a faster higher ionization zone (with $v_{\rm out}>1000$\,km\,s$^{-1}$), at least on the basis of the X-ray spectra from the present epoch of observations. 
Both the low velocity and ionization of the absorber in Mrk\,1040 likely places the gas at large distances from the black hole, as we will show below. 
This is far removed from some of the high velocity and high ionization components of the warm absorber discussed above, 
which may instead originate from an accretion disk wind on sub-parsec scales \citep[e.g.][]{tombesi2013} 
and some of which are powerful enough to drive gas out to kpc scales \citep{tombesi2015,feruglio2015}. 

\subsection{The Properties and Location of the Warm Absorber in Mrk 1040}

The best physical constraints on the absorbing gas in Mrk\,1040 arise from the zone 3 absorber (see Table 4), as this zone is also associated 
with the He-like emission from the Mg triplet, which makes it possible to calculate the 
covering fraction of the gas as well as its radial location. 
Photoionized emission spectra, in the form of an additive table of models (or an atable within \textsc{xspec}) were also generated with \textsc{xstar}, with the same properties as the 
zone\,3 absorber and with the same input continuum and ionization ($\log\xi=2.1$).

From the photoionization modeling, 
the normalization (or flux), $\kappa$, of an emission component is defined 
by \textsc{xstar} \citep{kallman2004} in terms of:
\begin{equation}
\kappa = f\frac{L_{38}}{D_{\rm kpc}^2}
\end{equation}
where $L_{38}$ is the ionizing luminosity in units of $10^{38}$\,erg\,s$^{-1}$ over the 1--1000\,Rydberg band and 
$D_{\rm kpc}$ is the distance to the AGN in kpc\footnote{see http://heasarc.gsfc.nasa.gov/docs/software/xstar/docs/html/node96.html}. Here $f$ is the covering fraction 
of the gas with respect to the total solid angle, where $f = \Omega / 4\pi$ and thus for a 
spherical shell of gas, $f=1$. 
For Mrk\,1040, where $L_{\rm ion}=1.8\times10^{44}$\,erg\,s$^{-1}$ and $D=68.7$\,Mpc \citep{theureau2007}, 
then for a spherical shell the expected {\sc xstar} normalization is $\kappa=3.8\times10^{-4}$. 
Hence for a given column density of gas, this sets the total luminosity of the soft X-ray photoionized emission; also see \cite{reeves2016} for a similar calculation.

We then applied this \textsc{xstar} emission component to the spectrum of Mrk\,1040 
in order to reproduce the He-like emission from Mg\,\textsc{xi}, which is the strongest emission line present in the 
spectrum and in particular the forbidden line component. Adopting the above normalization of the emission 
component, then a column density of $N_{\rm H}=1.0^{+0.2}_{-0.5}\times10^{21}$\,cm$^{-2}$ is required, 
in the case where the gas fully covers the X-ray source with a solid angle of $4\pi$\,steradian. Alternatively if we fix the column density of the emitter to what was previously measured for the
zone\,3 absorber (where $N_{\rm H}=1.5\times10^{21}$\,cm$^{-2}$, see Table\,4), then a minimum covering fraction 
of $f>0.4$ of $4\pi$\,steradian is required to model the emission.

A lower limit on the radial location of the emission can also be obtained from the upper limit to the velocity width 
of the line emission. Taking $3\sigma^2 = GM/R$ and with an upper limit for velocity width of the Mg\,\textsc{xi} forbidden emission of $\sigma_{\rm v}<135$\,km\,s$^{-1}$ (see Table\,3), then for an estimated black hole mass of Mrk\,1040 of 
$4\times10^{7}$\,M$_{\odot}$ \citep{zhou2010}, the location of the emitting gas associated with zone\,3 is 
found to be $>10^{19}$\,cm or $>3$\,pc. 
 
A typical limit to the radial location of the absorber can be derived
from the definition of the ionization parameter and the requirement
that the thickness of the absorber does not exceed its distance to the
supermassive black hole, i.e., $N_\mathrm{H} \simeq n_\mathrm{H} \Delta R <
n_\mathrm{H} R$. Then from the definition of the ionization parameter, this yields:-  
\begin{equation}
r_{\mathrm{max}} \equiv (L_{\mathrm{ion}}/\xi N_\mathrm{H}) (\Delta R/ R). 
\end{equation}
\noindent Thus for the condition that $\Delta R / R \sim 1$ and for the measured zone\,3 
absorber parameters of $\log\xi=2.1$, $N_{\rm H}=1.5\times10^{21}$\,cm$^{-2}$ and an
ionizing luminosity of $L_{\rm ion}=1.8\times10^{44}$\,erg\,s$^{-1}$, 
then $r_{\rm max}\sim 10^{21}$\,cm 
or $<300$\,pc. Note that the location of the gas cannot be much larger than this to satisfy the above 
geometric condition, while it can be smaller if $\Delta R/R < 1$ or if the gas is clumpy. 
 
For comparison, the expected sizescale of the Narrow Line Region (NLR) can be estimated by the 
scaling relation of \cite{Mor09}:-

\begin{equation} 
R_{\rm NLR} = 295 \times L_{46}^{0.47\pm 0.13} ~~~~ (pc).
\end{equation}

\noindent where here $L_{46}$ is the bolometric luminosity of the AGN in units of $10^{46}$\,erg\,s$^{-1}$.
If we take the 2-10\,keV luminosity to be a typical 3\% of the total bolometric luminosity of Mrk\,1040, 
then $L_{\rm bol}\sim7\times10^{44}$\,erg\,s$^{-1}$ and the estimated distance of the optical NLR 
would be $R_{\rm NLR}\sim100$\,pc, consistent with the above minimum and maximum estimates 
for the radial distance of the ionized gas (of $3\,{\rm pc}<R<300\,{\rm pc}$).
If the ionized X-ray gas is extended on distances of up to $\sim300$\,pc, 
then the density of the gas is low, of the order $n\sim1$\,cm$^{-3}$ for a 
column density of $N_{\rm H}=10^{21}$\,cm$^{-2}$. This may be expected for large scale gas associated 
with an extended NLR region. Note the distance scale of up to 300\,pc is consistent with archival 
HST images \citep{schmitt2003}, which show extended [O\,\textsc{iii}] emission within $\pm1$\arcs\ 
of the nucleus (see Section\,5.3).


We can place upper-limits to the mass outflow rate ($\dot{M}$), which in the quasi-spherical case is given by 
$\dot{M}=f4\pi N_{\rm H} R v_{\rm out} \mu m_{\rm p}$, where $f$ 
is the gas covering fraction and $\mu\sim1.2$ is the average particle mass relative to H. 
From the observations, $N_{\rm H}=1.5\times10^{21}$\,cm$^{-2}$, the covering fraction 
(as estimated from the emission) is $f\sim0.4$, while we adopt a likely upper-limit to the outflow velocity of $v_{\rm out}<100$\,km\,s$^{-1}$.
Thus for the most likely range in the radial location ($R$) of the absorber, i.e. from 3\,pc to 300\,pc as above, then the limits on the mass outflow rate 
are $\dot{M}<0.02$\,\Msun\,yr$^{-1}$ and $\dot{M}<2$\,\Msun\,yr$^{-1}$ respectively. Even for the most conservative upper-limit of $\dot{M}<2$\,\Msun\,yr$^{-1}$, 
then the corresponding upper-limit to the kinetic power of the outflow is very low with $L_{\rm K}<10^{40}$\,erg\,s$^{-1}$. 

Similar constraints can be placed on the other absorber zones. 
Given the lower ionizations of zones 1 and 2 (Table\,4), equation 2 
would predict $r_{\rm max}$ values 1--2 orders of magnitude greater than for the more highly ionized zone\,3 discussed above (i.e. on kpc scales). 
However a moderate clumping of the gas would restrict it to spatial scales consistent with the 
extended [O\,\textsc{iii}] emitting gas, which is discussed below.
Thus for the lowest ionization zone\,1 (with $\log\xi=0$), if $\Delta R/R \sim 0.02$ and for the measured column of
$N_{\rm H}=4\times10^{21}$\,cm$^{-2}$, then the distance is $r_{\max}\sim10^{21}$\,cm 
($\sim 300$\,pc) and thus the density would be $n \sim N_{\rm H}/\Delta R \sim 200$\,cm$^{-3}$. 
This is typical of the estimated densities for 
line emitting NLR clouds \citep{koski1978}. Furthermore, gas with this low ionization ($\log\xi\sim0)$ will be more than capable of producing [O\,\textsc{iii}] emission in the optical band. 
Indeed as will be shown from subsequent \xmmrgs\ soft X-ray spectroscopy of Mrk\,1040 (paper II), absorption is seen in the O K-shell band 
covering the full range of charges states; i.e. from O\,\textsc{i-viii}, including the strong detection of K-shell absorption from O\,\textsc{iii}. 
The latter absorption is also likely to be associated to the low ionization zone 1 gas.

\subsection{Imaging and Kinematics of the Nucleus with HST}

Interestingly, from the {\it HST} imaging of the [O\,\textsc{iii}] narrow line emitting gas in 
Mrk\,1040 \citep{schmitt2003}, the NLR appears to be extended on 
spatial scales of $\sim\pm1$\arcs\ from the nucleus. This is consistent with the above 
maximum estimate of $r_{\rm max}\sim300$\,pc for the warm absorber at the 
distance of Mrk\,1040. 
\citet{schmitt2003} report that the [O\,\textsc{iii}] emission appears bi-conical in structure, 
with an opening angle of 120\degg, with the cone axis approximately perpendicular to the 
host galaxy disk; the {\it HST} Wide Field Planetary Camera 2 (WFPC2) image of Mrk\,1040 
is reproduced in Figure\,12 (left panel) from these observations.
Note that no direct evidence is found for extended X-ray emission from the 
\chandra\ image of Mrk\,1040. However the brightness 
of the central AGN, the lower spatial resolution of \chandra\ and the moderate photon 
pile-up present in the zeroth order image precludes a quantitative comparison with the {\it HST} image. 

Two {\it HST} Space Telescope Imaging Spectrometer (STIS) 
observations of Mrk\,1040 in 2009 were performed with the G430M grating, sampling 
two different position angles through the central part of the galaxy. The details of the observations 
and of the analysis proceedures are reported by \cite{fischer2013}, who perform imaging and kinematics 
(primarily in [O\,\textsc{iii}]) of a sample nearby Seyfert galaxies with {\it HST} STIS.
Here we show the results obtained in Mrk\,1040, in order to compare the location and 
kinematics of the narrow line region emitting gas with the X-ray warm absorber. 

These results are summarized in Figure\,12. The two STIS observations give two slit positions 
through different portions of the galaxy, with slit A along the major axis of the galaxy and 
slit B approximately perpendicular to this. The positions of the two slits are indicated 
in the left hand panel of Figure\,12, overlaid on the non dispersed WFPC2 image 
of Mrk\,1040 at [O\,\textsc{iii}] $\lambda 5007$. The center and right panels of Figure\,12 
show the resulting 
kinematics of the [O\,\textsc{iii}] line components versus position in arcseconds from the 
AGN center. These plots indicate that there is evidence for some modest outflow located within 
$\pm0.2$\arcs\ of the nucleus of Mrk\,1040, with outflow velocities of typically $-200$\,km\,s$^{-1}$.
This blue-shifted component of [O\,\textsc{iii}] is visible in both of the slit A and B spectra 
and occurs well within the inner 100\,pc of the central AGN.
Slit A, positioned along the major axis, also shows extended, blue-shifted [O\,\textsc{iii}] emission 
up to 0.5\arcs\ to the southwest of the nucleus and in the direction of the major axis of the galaxy. 
This emission may be attributed to rotation, if for instance 
the disk is rotating in a counterclockwise direction and thus the southwest side 
would appear blue shifted. 

\subsection{Implications for the X-ray Absorber}

Thus the likely distance of the ionized X-ray gas 
is consistent with the extended emission observed from the optical NLR, 
as seen in the HST observations. The X-ray kinematics are also broadly consistent with what is 
seen in {\it HST}, where some modest outflow is associated with the nucleus itself of up to 
$-200$\,km\,s$^{-1}$. Indeed for the zone 1 absorber, which is most likely associated to the 
[O\,\textsc{iii}] emission, the outflow velocity is $-150^{+105}_{-100}$\,km\,s$^{-1}$ (see Table 4).
The aperture of \chandra\ HETG is much larger than for the {\it HST} 
STIS and thus given the likely upper limit of $<300$\,pc to the location of the X-ray absorber, 
we cannot exclude either an origin either associated with the weak nuclear outflow located 
within $<100$\,pc of the nucleus as seen in [O\,\textsc{iii}], or with the somewhat more extended 
gas associated with the rotating galactic disk, with low radial velocities.
A tighter constraint on the radial distance of the X-ray absorbing gas would be needed 
to distinguish between these possibilities; this will be discussed further in paper II, where 
we present evidence for variations in absorber 
ionization between the different epochs. 

One interesting scenario is that we are viewing a failed nuclear wind from Mrk\,1040. 
The accretion rate of Mrk\,1040, given its luminosity and likely black hole mass, 
probably does not exceed 10\% of Eddington. With such a weak AGN, 
it may be difficult to radiatively accelerate gas at distances beyond $\sim100$\,pc 
\citep{fischer2016}, although this gas could still be ionized by the AGN.
Indeed the stringent upper limit obtained earlier on the outflow kinetic power 
($L_{\rm K}<10^{40}$\,erg\,s$^{-1}$) implies that this weak wind will leave a minimal
kinetic footprint on its host galaxy, while the X-ray outflow velocity is also no larger than 
the velocity dispersion of the stellar bulge in Mrk\,1040 \citep{nelson&whittle1995}.
The low X-ray velocity may also be in part due to our orientation, if for instance 
we are viewing Mrk\,1040 pole-on compared to a more equatorial wind.
The narrowness of the Balmer lines in Mrk\,1040 is more likely 
due to a pole-on view, rather than a high Eddington rate as is usually inferred in more typical NLS1s. 

\section{Conclusions}

This paper has presented a 200\,ks \chandra\ HETG 
observation of the nearby Seyfert 1, Mrk\,1040, which is the first time a high resolution 
X-ray spectrum of this AGN has been discussed. The spectra show the strong signature of a warm absorber, 
with the lower ionization gas responsible for the strong series of inner K-shell absorption lines from Ne, Mg and Si, 
originating from ions with low charge states. 
Neither the low or high ionization components of the absorber appear to be outflowing, to typically within 
$\pm100$\,km\,s$^{-1}$ of the systemic velocity of Mrk\,1040. 
We also find that the lower ionization (inner K shell) absorption lines are 
best reproduced with a model that requires a steep ($\Gamma=2.5$) 
photon index below 1\,keV. This argues for an intrinsic origin for the 
soft X-ray excess from this AGN.

The low ionization and low velocity of the absorber 
likely places the absorbing gas between $\sim3$ and $\sim300$\,pc from the black hole and it appears coincident with the optical NLR emitting gas, 
as seen in [O\,\textsc{iii}] in the {\it HST} images of this AGN.
This is also broadly similar to the soft X-ray absorbers seen towards other 
Seyfert 1s, which may typically range from a pc scale toroidal wind \citep{krolik&kriss2001,blustin2005}, to the 
NLR on scales of several tens to hundreds of parsecs \citep[e.g.][]{netzer2003,behar2003,kriss2011,kaastra2012}. 
Finally we suggest that the lack of an X-ray outflow component  
in Mrk\,1040 may be due to a failed nuclear wind in this AGN, with the gas not able to 
accelerate out to scales beyond $\sim100$\,pc.

\section{Acknowledgements}

We thank the anonymous referee for their positive and constructive comments.
J.N.\ Reeves acknowledges Chandra grant number GO3-14123X, as well as NASA grant 
numbers NNX16AE11G and NNX15AF12G. T.J.\ Turner acknowledges NASA grant number NNH13CH63C.
Both J.N.\ Reeves and A.\ Lobban acknowedge support from STFC, 
via the consolidated grants ST/M001040/1 and ST/K001000/1.
D. Porquet acknowledges financial support from the European Union Seventh Framework Program 
(FP7/2007-2013) under grant agreement number 312789. 
E.\ Behar is supported by the the European Union’s Horizon 2020 research and innovation programme under the Marie Sklodowska-Curie grant agreement no. 655324 and by the I-CORE program of the Planning and Budgeting Committee (grant number 1937/12).
E.N also acknowledges funding from the European Union’s Horizon 2020 research and innovation programme under the Marie Skłodowska-Curie grant agreement No. 664931.
TCF was supported by an appointment to the NASA Postdoctoral Program at the NASA Goddard Space Flight Center, administered by Universities Space Research Association under contract with NASA.
The scientific results reported in this article are based on observations made by the Chandra X-ray Observatory.
This research has made use of software provided by the Chandra X-ray Center (CXC) in the application package CIAO.

\clearpage

\clearpage

\clearpage

\clearpage

\begin{figure}
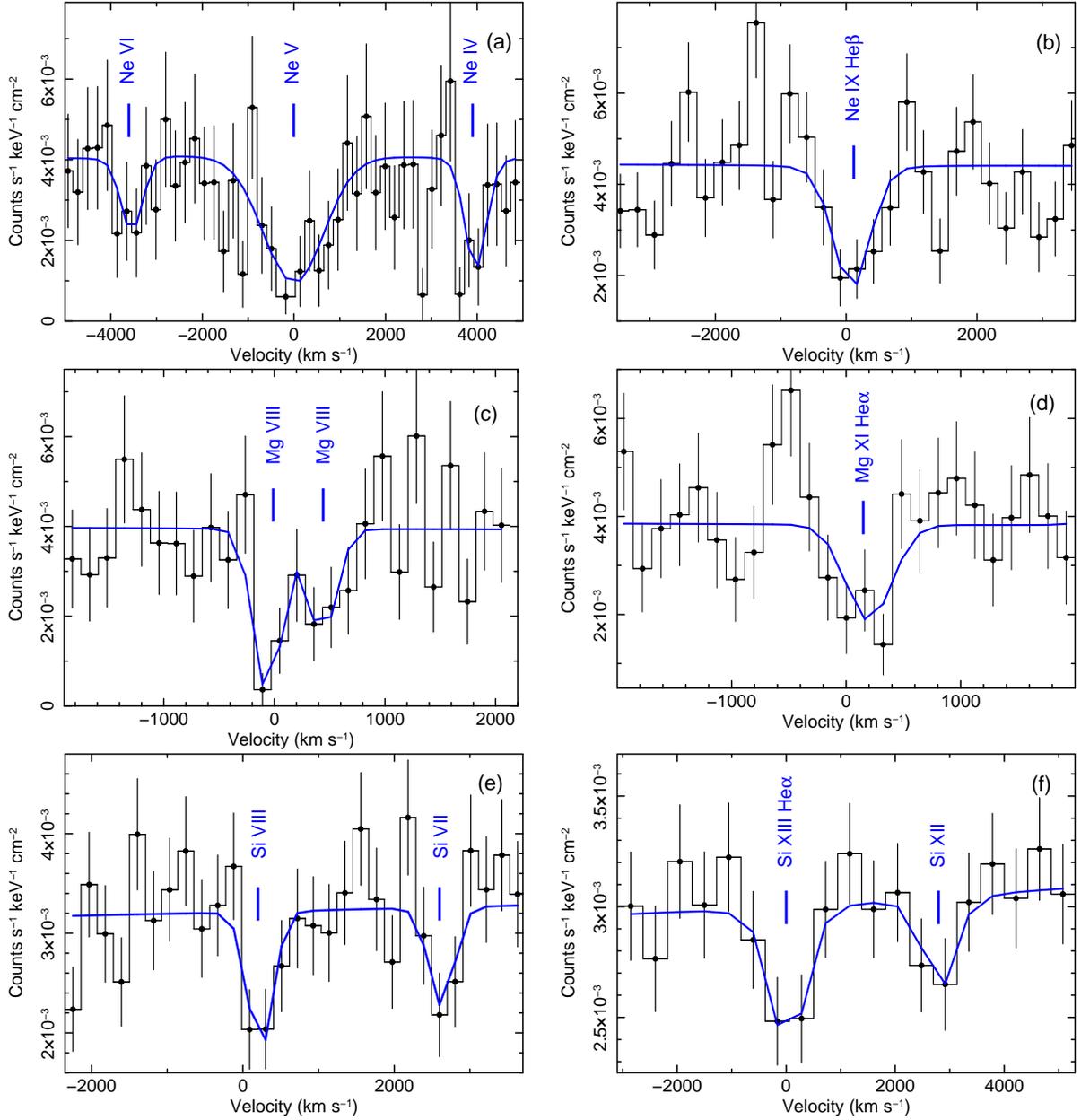

\begin{center}
\includegraphics[angle=-90,width=0.49\textwidth]{f4a.eps}
\includegraphics[angle=-90,width=0.49\textwidth]{f4b.eps}
\includegraphics[angle=-90,width=0.49\textwidth]{f4c.eps}
\includegraphics[angle=-90,width=0.49\textwidth]{f4d.eps}
\includegraphics[angle=-90,width=0.49\textwidth]{f4e.eps}
\includegraphics[angle=-90,width=0.49\textwidth]{f4f.eps}
\end{center}
\caption{Velocity profiles of selected He-like and inner-shell 
absorption lines corresponding to, (a) Ne\,\textsc{v}, (b) 
Ne\,\textsc{ix} (He-$\beta$), (c) Mg\,\textsc{viii}, (d) Mg\,\textsc{xi}, (e) Si\,\textsc{viii} and 
(f) S\,\textsc{xiii}. The profiles for Ne are extracted from the MEG, while the other lines are extracted 
from the HEG, at HWHM resolution.
Note negative velocities denote blue-shift 
compared to the expected rest frame energies listed in Table\,2. The solid blue lines show the best-fit 
Gaussian profiles. Other nearby lines from adjacent ions are marked on the plots, while the Mg\,\textsc{viii}
absorption is from a resolved doublet centered on the stronger of the two lines.
Generally, the absorption lines appear centered close 
to zero velocity (indicating no outflow), with their best fit line widths and velocity shifts reported in Table 2. 
See text for a detailed discussion.}
\label{profiles}
\end{figure}
\clearpage

\begin{figure}
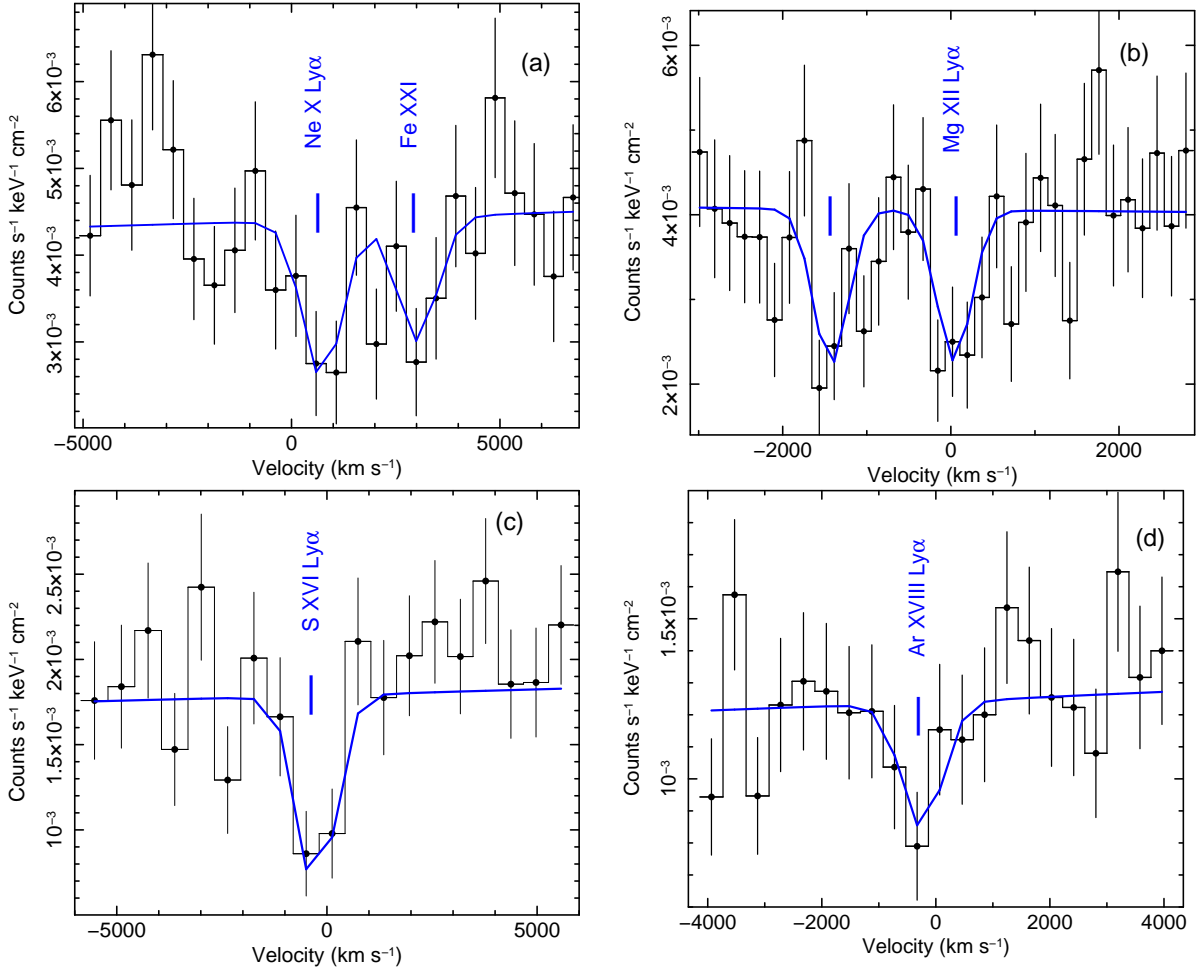

\begin{center}
\includegraphics[angle=-90,width=0.49\textwidth]{f5a.eps}
\includegraphics[angle=-90,width=0.49\textwidth]{f5b.eps}
\includegraphics[angle=-90,width=0.49\textwidth]{f5c.eps}
\includegraphics[angle=-90,width=0.49\textwidth]{f5d.eps}
\end{center}
\caption{As per Figure\,4, but showing the velocity profiles of selected H-like ($1s-2p$) absorption lines corresponding to, (a) Ne\,\textsc{x} (He-$\alpha$), (b) Mg\,\textsc{xii}, (c) S\,\textsc{xvi} and (d) 
Ar\,\textsc{xviii}. 
The profiles have been extracted from the HEG, with the exception of Ne\,\textsc{x} (MEG). 
As per the other profiles, the H-like line profiles generally appear to be unresolved, 
with outflow velocities consistent with zero (see Table\,2). The Mg\,\textsc{xii} profile does appear to show evidence for a 2nd, blue-shifted component at $-1400$\,km\,s$^{-1}$, however this does not appear to be present in the other H-like profiles and this component is not confirmed in the MEG data. 
Note that the Ne\,\textsc{x} He-$\alpha$ profile contains a contribution 
from Fe\,\textsc{xxi} absorption red-wards of the absorption line profile and the velocity profile 
is better determined in the Lyman-$\beta$ line profile which is uncontaminated.}
\label{profiles2}
\end{figure}
\clearpage

\begin{figure}
\begin{center}
\rotatebox{-90}{\includegraphics[width=11cm]{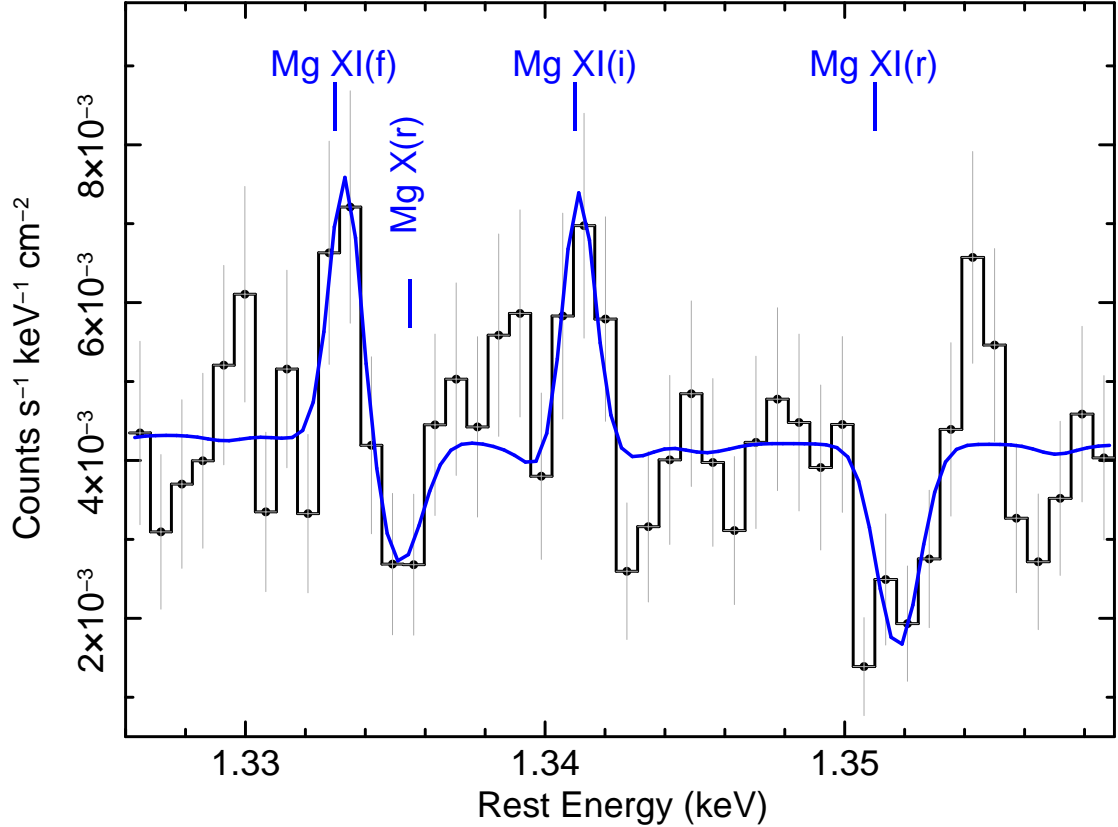}}
\end{center}
\caption{Zoom in around the Mg\,\textsc{xi} He-like triplet for the HEG spectrum, binned to 
HWHM resolution. Count rate data are 
shown in black, the model is shown in blue. Excess emission is present near the expected position of the 
forbidden and intercombination lines, as is marked on the plot, while absorption is clearly observed 
at the position of the resonance transition. The emission lines are unresolved with a velocity width of 
$\sigma<135$\,km\,s$^{-1}$. Note the weak absorption immediately bluewards of the forbidden emission line 
is due to the Li-like resonance ($1s\rightarrow2p$) line of Mg\,\textsc{x}.} 
\label{Mg}
\end{figure}

\clearpage

\begin{figure}
\begin{center}
\rotatebox{-90}{\includegraphics[width=11cm]{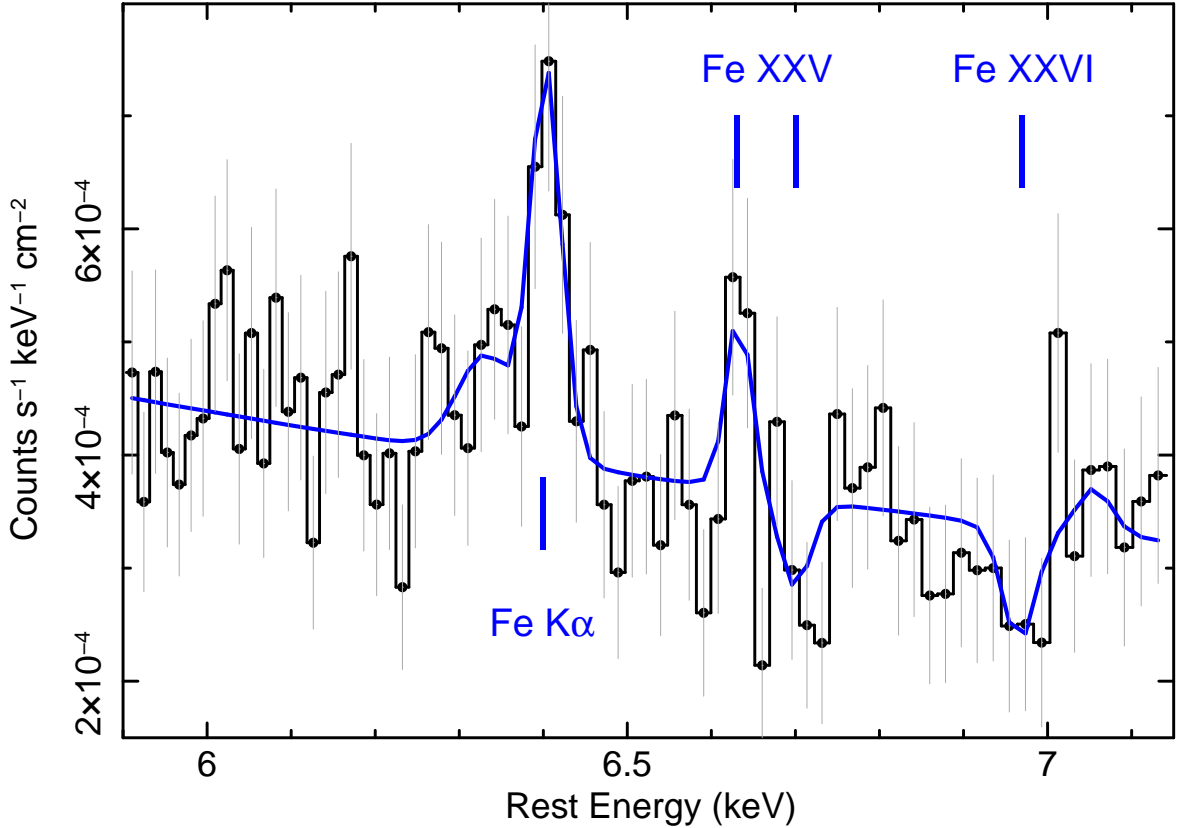}}
\end{center}
\caption{Iron K band ratio residuals to the HEG spectrum of Mrk\,1040 (at HWHM resolution), 
compared to the best fit continuum and warm absorber model. The narrow iron K$\alpha$ line at 6.4\,keV, 
with an equivalent width of 44\,eV, is clearly apparent and is unresolved with a upper-limit to its line width of $\sigma<28$\,eV 
(or $<1300$\,km\,s$^{-1}$). A weak Compton shoulder (on the red-wing of the K$\alpha$ line) may also be present and the profile 
is consistent with reflection off distant, near neutral matter.  Emission from the forbidden 
line of Fe\,\textsc{xxv} is present and is unresolved with a line width of $\sigma<14$\,eV 
(or $<630$\,km\,s$^{-1}$) and also likely originates from distant matter. The line contribution of the warm absorber to the iron K profile is relatively modest, with some weak absorption predicted 
from the resonance lines of He and H-like Fe. 
Note that after accounting for the continuum curvature due to the warm absorber, 
no residuals from a broad emission line are apparent in the Fe K band.} 
\label{fe}
\end{figure}

\clearpage

\begin{figure}
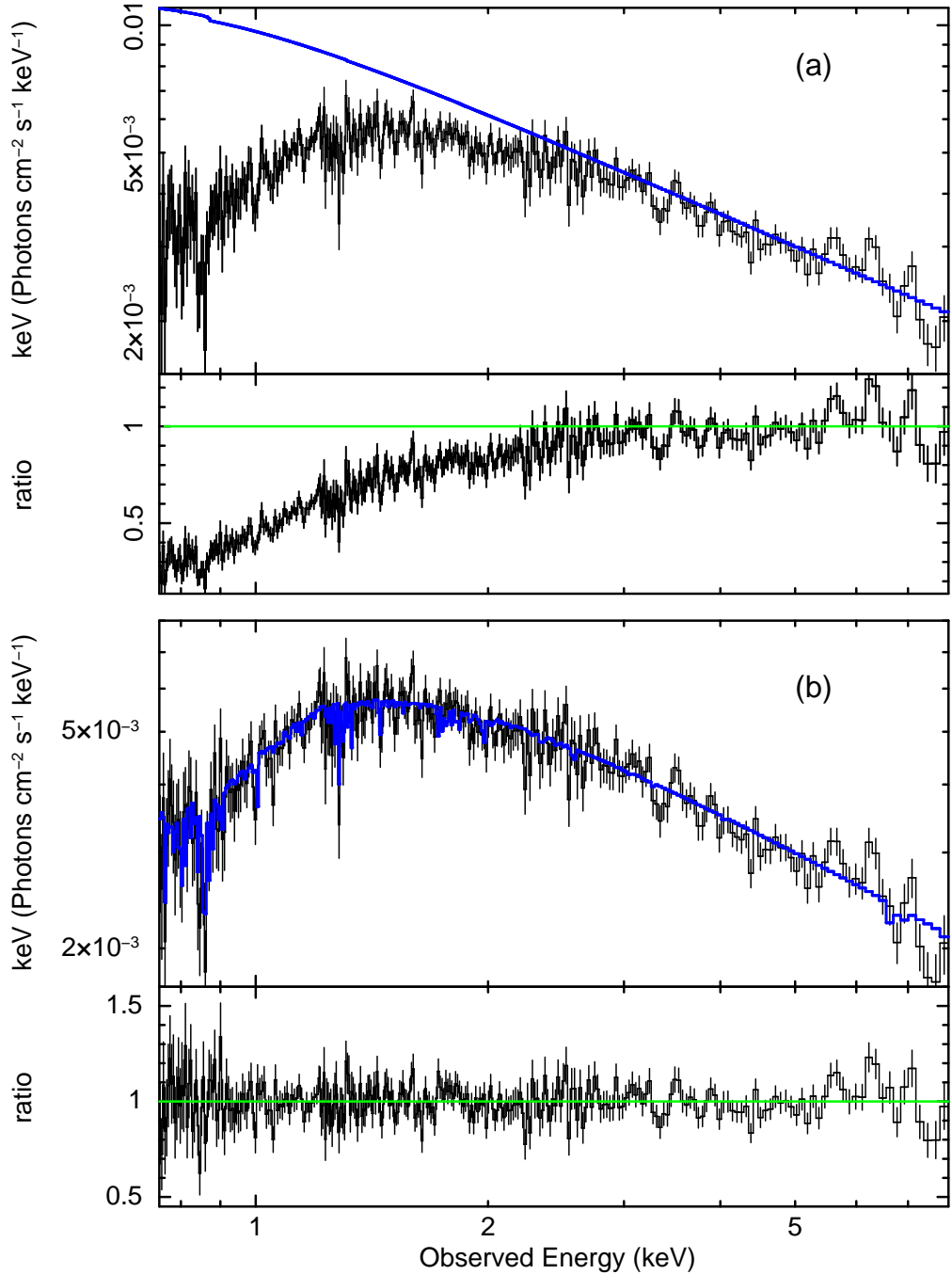

\begin{center}
\rotatebox{-90}{\includegraphics[height=14cm]{f8a.eps}}
\rotatebox{-90}{\includegraphics[height=14cm]{f8b.eps}}
\end{center}
\caption{Broad band view of the Chandra HETG spectrum, binned by a factor $\times 2$ the FWHM resolution for clarity. In the upper (a) panel the fluxed spectrum is compared to a simple power-law continuum of $\Gamma=1.8$ fitted above 3\,keV and attenuated only the Galactic absorption column. The curvature towards lower energies due to the warm absorber is apparent in the data/model ratio residuals. The lower panel (b) shows the spectrum with the best-fit warm absorber 
model overlaid. In this case the warm absorber model is able to account for the spectral curvature towards 
soft X-ray energies. Note that the emission lines have not been included at Mg or Fe K (due to the coarser binning adopted), these are shown in the previous Figures\,6 and 7.}
\label{spectra}
\end{figure}

\clearpage

\begin{figure}
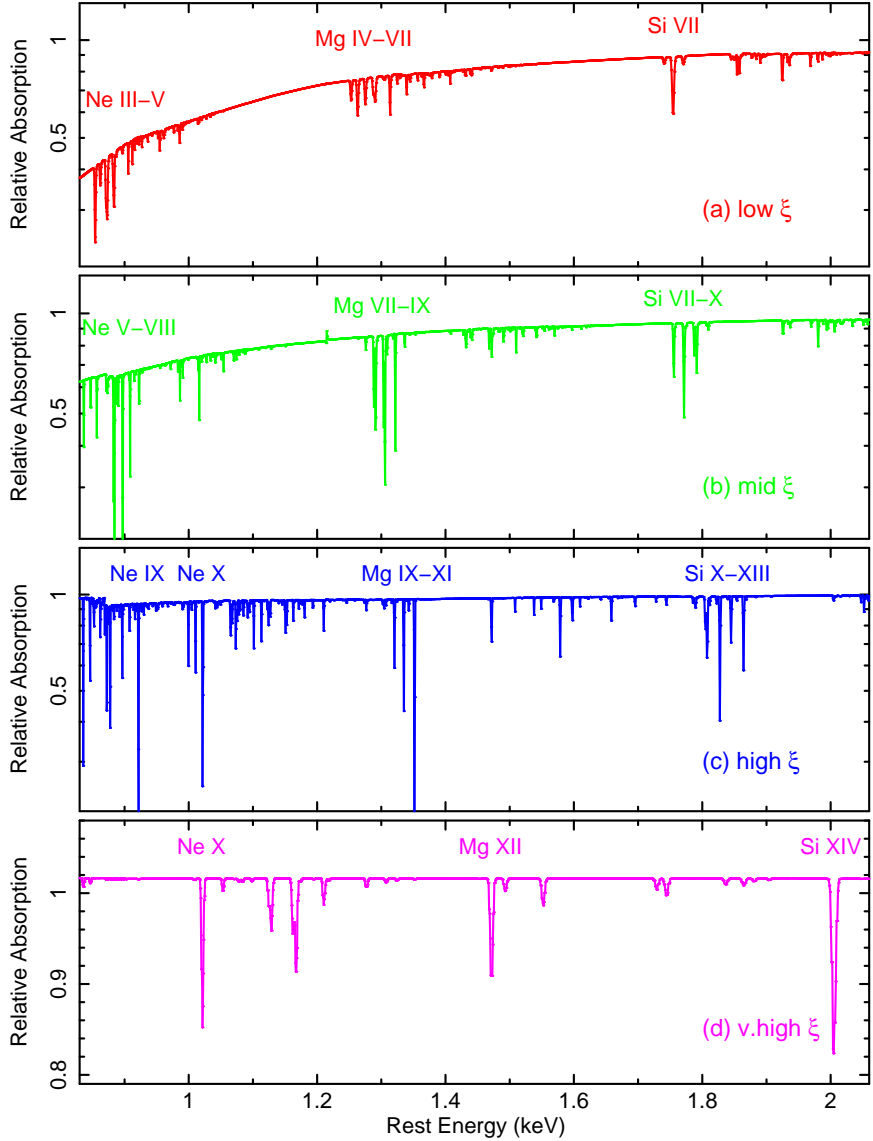

\centering
\rotatebox{-90}{\includegraphics[height=12cm]{f9a.eps}}
\rotatebox{-90}{\includegraphics[height=12cm]{f9b.eps}}
\rotatebox{-90}{\includegraphics[height=12cm]{f9c.eps}}
\rotatebox{-90}{\includegraphics[height=12cm]{f9d.eps}}
\caption{Relative contribution of respective warm absorption zones towards the attenuation of the 
soft X-ray spectrum. 
The four absorption components, from lowest to highest ionization, correspond to the warm absorber zones listed in Table\,4.
The lower ionization zones 1 and 2 (panels a and b)
have the largest opacity, with absorption due to inner shell Ne, Mg, Si; 
these zones are responsible for much of the soft X-ray spectral curvature and absorption line 
structure. The higher ionization zone 3 (panel c) 
contains absorption mainly due to Li and He-like ions, whilst the highest ionization zone (panel d)  
has the lowest opacity and contributes only to the H-like lines.}
\label{components}
\end{figure}

\clearpage

\begin{figure}
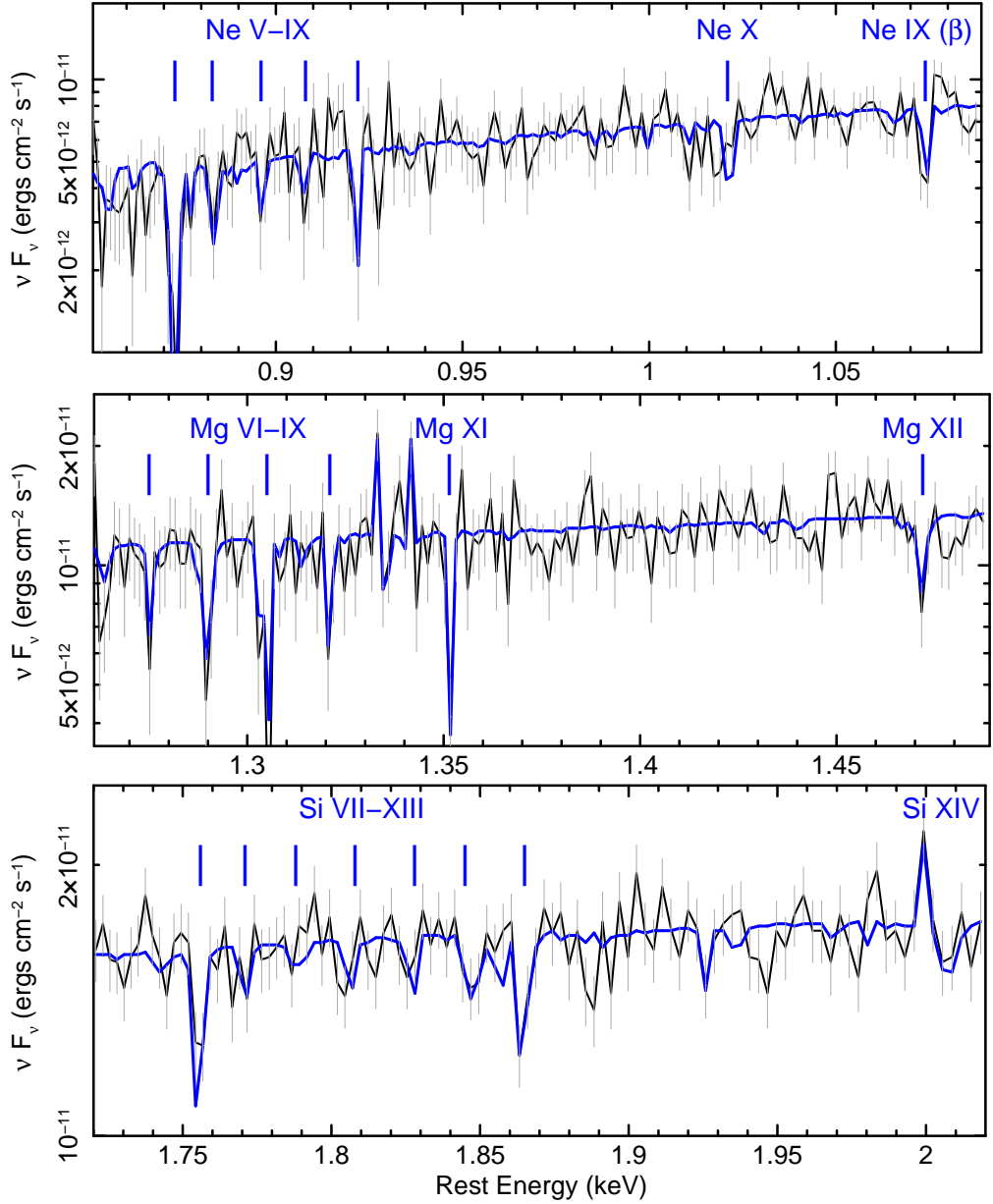

\begin{center}
\rotatebox{-90}{\includegraphics[height=14cm]{f10a.eps}}
\rotatebox{-90}{\includegraphics[height=14cm]{f10b.eps}}
\rotatebox{-90}{\includegraphics[height=14cm]{f10c.eps}}
\end{center}
\caption{The warm absorber model, shown in blue, superimposed on the HETG spectrum binned at FWHM 
resolution. The upper panel shows the MEG data and the lower two panels the HEG data. A three zone 
warm absorber with ionization in the range $\log(\xi/{\rm erg\, cm\, s}^{-1})=0-2$ is able to account 
for the 
absorption in both the Ne (upper panel), Mg (middle panel) and Si (lower panel) bands, 
covering the lower ionization inner-shell absorption lines to the highly ionized He/H-like lines. 
The warm absorber parameters are 
tabulated in Table\,4 and the net velocities of the zones appear consistent with zero, suggesting that 
none of the zones appear to be associated with outflowing gas.}
\label{xstar}
\end{figure}

\begin{figure}
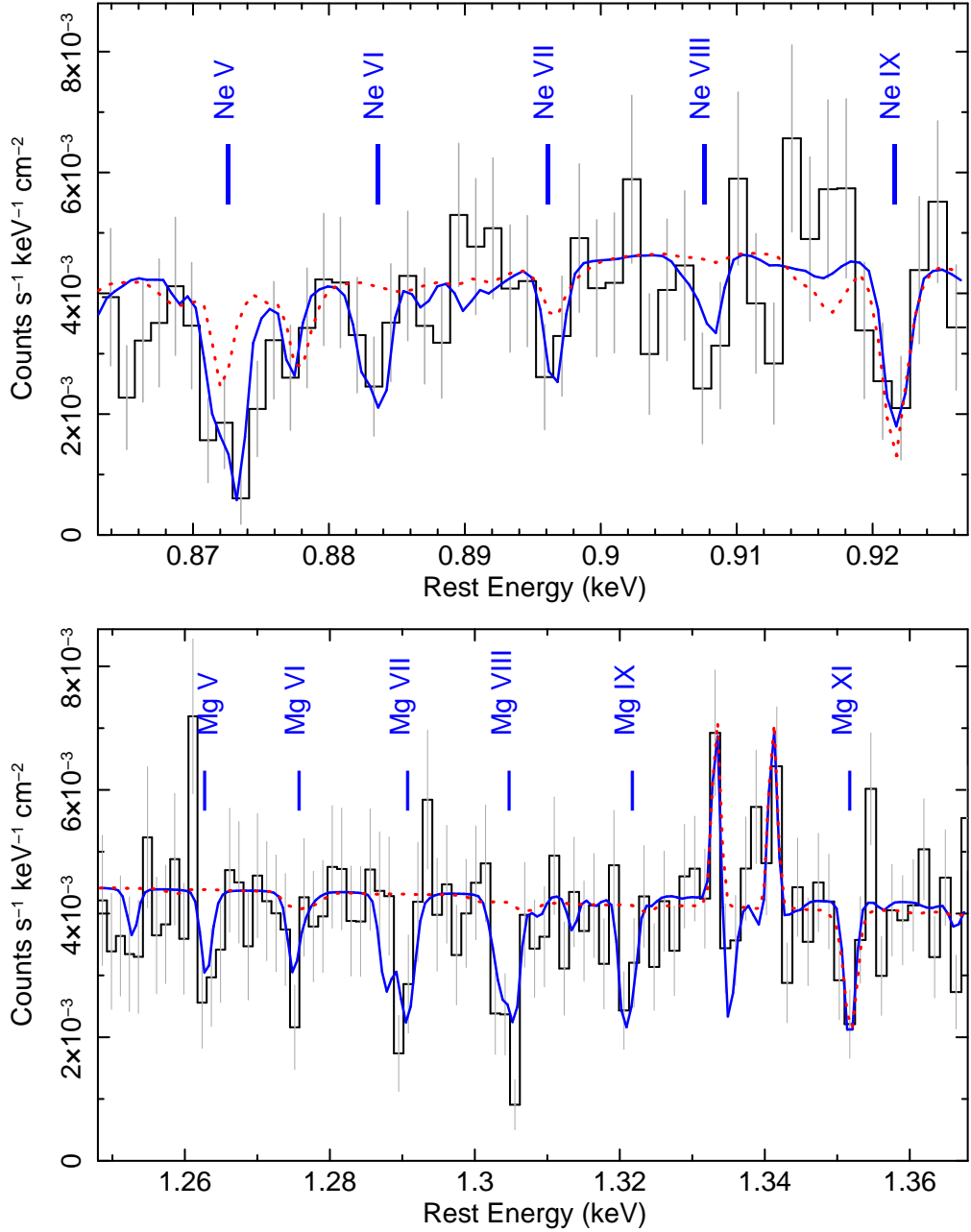

\begin{center}
\rotatebox{-90}{\includegraphics[height=14cm]{f11a.eps}}
\rotatebox{-90}{\includegraphics[height=14cm]{f11b.eps}}
\end{center}
\caption{Comparison between warm absorber models in the Ne (upper panel) and 
Mg (lower panel) K-shell bands. The solid blue line shows the best fit absorber model as 
described in Section 4.2, 
which assumes a steep UV to soft X-ray photoionizing continuum of $\Gamma=2.5$ and 
is able to reproduce all of the absorption features shown. On the other hand the dotted 
red line shows an alternative absorber model, where a flat $\Gamma=2$ photoionizing continuum 
has been assumed from $1-1000$\,Rydberg. This latter model is unable to reproduce 
the depths of the low ionization inner shell absorption lines that are apparent 
from Ne\,\textsc{v-viii} and Mg\,\textsc{vi-ix} and can only account for high ionization 
lines, such as Ne\,\textsc{ix} or Mg\,\textsc{xi}. Thus the warm absorber models are 
sensitive to the assumed slope of the intrinsic soft X-ray continuum. Note the 
spectra have been binned to the FWHM resolution of the MEG and HEG respectively.}
\label{continuum}
\end{figure}

\begin{figure}
\begin{center}
\rotatebox{0}{\includegraphics[width=17cm]{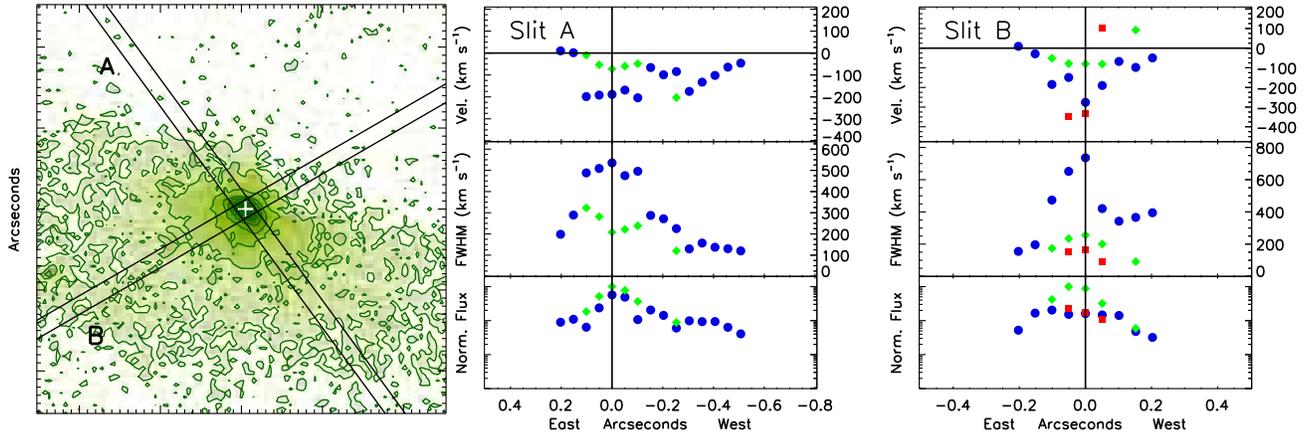}}
\end{center}
\caption{HST imaging observations of the nucleus of Mrk\,1040. The left hand plot shows the 
WFPC2 image of Mrk\,1040 at [O\,\textsc{iii}] $\lambda 5007$ showing the central $2.5\times2.5$\arcs\ 
region of the galaxy centered on the AGN (white cross). North is up, east is to the left. 
At the distance of Mrk\,1040 the scale is 330\,pc\,arcs$^{-1}$. 
The position of the two slits used for the 
HST STIS observations (with the G430M grating) are marked A and B respectively. Slit A is positioned 
along the major axis of the galaxy. 
The middle and right panels show the results of the kinematics obtained from the 
[O\,\textsc{iii}] line fitting for the two slit positions versus the position (in arcseconds) 
from the AGN center. 
The top panels shows the velocity shift of the [O\,\textsc{iii}] emission vs. 
position, where negative values 
denote blueshift. Thus a small component of outflow of $\sim-200$\,km\,s$^{-1}$ is found 
within $\pm0.2$\arcs\ of the nucleus in both of the slit A and B spectra, while a 
small blueshifted component is seen up to 0.5\arcs\ SW of the nucleus for slit A only -- 
this latter component may be due to rotation.
The middle panels show the FWHM of the [O\,\textsc{iii}] line components fitted, while 
the lower panel shows the relative [O\,\textsc{iii}] flux vs. position. 
The different colored points illustrate the kinematics of different components of [O\,\textsc{iii}]
selected on FWHM.
Thus blue, green, and red points signify the widest, second widest, and third widest components respectively, as can be seen from their FWHM values in the middle panels.}
\label{hst}
\end{figure}

\clearpage

\begin{deluxetable}{lcccc}
\tabletypesize{\small}
\tablecaption{Summary of Mrk\,1040 Chandra Observations}
\tablewidth{0pt}
\tablehead{
\colhead{Obsid} & \colhead{Start Date/Time$^{a}$} & \colhead{Inst} 
& \colhead{Exposure (ks)} & \colhead{Net Rate s$^{-1}$}}

\startdata
15075 & 2014-02-25 04:00:17 & HEG & 28.6 & $0.231\pm0.004$ \\
& -- & MEG & -- & $0.432\pm0.004$ \\
15076 & 2013-10-19 09:24:53 & HEG & 19.7 & $0.237\pm0.005$ \\
& -- & MEG & -- & $0.444\pm0.005$ \\
16571 & 2014-03-03 12:26:42 & HEG & 89.6 & $0.240\pm0.002$ \\
& -- & MEG & -- & $0.453\pm0.002$\\
16584 & 2014-02-26 01:17:35 & HEG & 60.0 & $0.247\pm0.002$ \\
& -- & MEG & -- & $0.459\pm0.003$\\
\hline
Total & -- & HEG & 197.8 & $0.238\pm0.001$ \\
& -- & MEG & 197.8 & $0.457\pm0.002$ \\
\enddata

\tablenotetext{a}{Observation Start/End times are in UT.} 
\label{observations}
\end{deluxetable}

\clearpage

\begin{deluxetable}{lcccccc}
\tabletypesize{\small}
\tablecaption{Gaussian Profiles of Prominent Soft X-ray Absorption Lines}
\tablewidth{0pt}
\tablehead{
\colhead{Line ID} & \colhead{$E_{\rm rest}$$^{a}$} & \colhead{$E_{\rm lab}$$^{a}$} & \colhead{$v_{\rm out}$$^{b}$}
& \colhead{$\sigma_{\rm width}$$^{c}$} 
& \colhead{EW(eV)}& \colhead{$\Delta C$$^d$}} 
\startdata
Ne\,\textsc{iv} (N-like) & $861.8\pm0.6$ & $862.7$ & $+310\pm210$ & $170^{+140}_{-120}$ & $-1.3\pm0.5$ & 8.1 \\
Ne\,\textsc{v} (C-like) & $873.1\pm0.7$ & $873.0$ & $-30\pm240$ & $490\pm200$ & $-3.8^{+0.9}_{-0.7}$ & 30.3 \\
Ne\,\textsc{vi} (B-like) & $883.3^{+1.0}_{-0.8}$ & $883.4$ & $+30^{+270}_{-340}$ & $170^{t}$ & $-0.9\pm0.5$ & 6.7 \\
Ne\,\textsc{ix} He-$\alpha$ & $921.5^{+0.6}_{-0.8}$ & $922.0$ & $+160^{+260}_{-195}$ & $<520$ & $-1.4^{+0.8}_{-0.5}$ & 13.0 \\
Ne\,\textsc{ix} He-$\beta$ & $1073.4\pm0.6$ & 1073.7 & $+80\pm170$ & $225^{+140}_{-100}$ & 
$-1.6^{+0.5}_{-0.4}$ & 21.2 \\
Ne\,\textsc{x} Ly-$\alpha$$^{f}$ & $1018.5^{+2.7}_{-0.8}$ & $1021.5$ & -- & -- & 
$-1.7\pm0.5$ & 11.5 \\
Ne\,\textsc{x} Ly-$\beta$ & $1211.0^{+0.6}_{-0.4}$ & $1210.9$ & $-25^{+100}_{-150}$ & $<270$ & 
$-0.7\pm0.3$ & 9.6 \\
Mg\,\textsc{v} (O-like) & $1264.2\pm0.4$ & $1263.1$ & $-260\pm100$ & $<100$ & 
$-0.8\pm0.4$ & 6.1 \\
Mg\,\textsc{vi} (N-like) & $1274.9\pm0.4$ & $1275.8$ & $+210\pm100$ & $<100^t$ & 
$-0.9\pm0.4$ & 6.5 \\
Mg\,\textsc{vii} (C-like) & $1289.4^{+0.4}_{-0.3}$ & $1289.0$ & $-90^{+70}_{-90}$ & $<280$ & 
$-0.9\pm0.4$ & 7.1 \\
Mg\,\textsc{viii} (B-like)$^{e}$ & $1305.6\pm0.2$ & $1305.4$ & $-50^{+45}_{-40}$ & $<105$ & $-1.5^{+0.4}_{-0.2}$ & 20.7 \\
Mg\,\textsc{viii} (B-like)$^{e}$ & $1303.5^{+0.5}_{-0.7}$ & $1303.2$ & $-70^{+160}_{-115}$ & $105^t$ & $-1.0^{+0.5}_{-0.4}$ & 7.7 \\
Mg\,\textsc{xi} He-$\alpha$ & $1351.7^{+0.4}_{-0.7}$ & $1352.2$ & $+110^{+200}_{-90}$ & $150^{+110}_{-95}$ & $-1.1^{+0.5}_{-0.4}$ & 19.8 \\
Mg\,\textsc{xii} Ly-$\alpha$ & $1472.0^{+0.6}_{-0.8}$ & $1472.2$ & $+40^{+160}_{-120}$ & $150^{+140}_{-90}$ & $-1.1^{+0.4}_{-0.5}$ & 14.6 \\
Si\,\textsc{vii} (O-like) & $1756.1^{+0.6}_{-1.1}$ & $1755.4$ & $-120^{+190}_{-120}$ & $<150$ & $-0.9\pm0.5$ & 7.8 \\
Si\,\textsc{viii} (N-like) & $1770.3^{+0.5}_{-0.7}$ & $1771.4$ & $+185^{+120}_{-85}$ & $<150^t$ & $-1.1^{+0.5}_{-0.4}$ & 12.6 \\
Si\,\textsc{xi} (Be-like) & $1827.2^{+1.2}_{-1.0}$ & $1829.2$ & $+330^{+160}_{-120}$ & $<210$ & $-0.9\pm0.4$ & 6.4 \\
Si\,\textsc{xii} (Li-like) & $1847.1\pm0.9$ & $1845.5$ & $-260\pm150$ & $<210^t$ & $-0.7\pm0.4$ & 6.3 \\
Si\,\textsc{xiii} He-$\alpha$ & $1865.2^{+0.8}_{-0.7}$ & $1865.0$ & $-30^{+110}_{-130}$ & $<210^t$ & $-1.2\pm0.5$ & 12.4 \\
S\,\textsc{xvi} Ly-$\alpha$ & $2624.1^{+2.1}_{-1.9}$ & $2622.0$ & $-240^{+220}_{-240}$ & $<520$ & $-5.4^{+2.1}_{-1.7}$ & 22.5 \\
Ar\,\textsc{xviii} Ly-$\alpha$ & $3322.6\pm1.6$ & $3320.6$ & $-180\pm150$ & $<950$ & $-3.4^{+2.3}_{-1.7}$ & 8.5 \\
\enddata
\tablenotetext{a}{$E_{\rm rest}$ is the measured line centroid (units eV), in the AGN rest frame. $E_{\rm lab}$ denotes the expected lab-frame energy, which for the He or H-like lines are taken from www.nist.gov, while values for the inner shell lines (O through to Li-like ions) are taken from Behar \& Netzer (2002). Where a line is a (unresolved) doublet, the weighted mean of the expected lab energies is calculated.}
\tablenotetext{b}{Velocity shift of absorption line in km\,s$^{-1}$. Negative 
values denote blue-shift, positive values redshift, with the uncertainty range given at 90\% confidence.}
\tablenotetext{c}{Intrinsic $1\sigma$ velocity width of absorption line in km\,s$^{-1}$.}
\tablenotetext{d}{Improvement in fit statistic after modeling Gaussian absorption profile.}
\tablenotetext{e}{Resolved absorption line doublet modeled by two Gaussian profiles.}
\tablenotetext{f}{The velocity shift and width of the Ne\,\textsc{x} Lyman-$\alpha$ is poorly 
determined, due to the presence of the nearby Fe\,\textsc{xxi} absorption line.}
\tablenotetext{t}{Velocity width of line tied to the best-fit value from an adjacent line.}
\label{line-widths}
\end{deluxetable}

\clearpage

\begin{deluxetable}{lccccc}
\tabletypesize{\small}
\tablecaption{X-ray Emission Lines in the Chandra HETG Spectrum of Mrk\,1040}
\tablewidth{0pt}
\tablehead{
\colhead{Line} & \colhead{E$_{\rm quasar}^{a}$} & \colhead{Flux$^{b}$} & \colhead{EW$^{c}$} 
& \colhead{$\sigma_{\rm v}$$^{d}$} &  \colhead{$\Delta C$$^{e}$}}
\startdata
Mg\,\textsc{xi} $(f)$ & $1333.3\pm0.5$ & $0.70^{+0.45}_{-0.40}$ & $1.1\pm0.7$ & $<135$ & 
9.7 \\
Mg\,\textsc{xi} $(i)$ & $1341.2\pm0.5$ & $0.70^{+0.48}_{-0.39}$ & $1.1\pm0.7$ & $<135$$^t$ & 
9.7 \\
Si\,\textsc{xiv} Ly-$\alpha$ & $1999^{+15}_{-8}$ & $0.42^{+0.26}_{-0.24}$ & $1.3\pm0.8$ & $<390$ & 
8.3 \\
Fe K$\alpha$ & $6403^{+12}_{-8}$ & $1.8^{+0.7}_{-0.5}$ & $44^{+17}_{-12}$ & $<1310$ & 
35.2 \\
Fe\,\textsc{xxv} $(f)$ & $6630^{+13}_{-6}$ & $0.70^{+0.45}_{-0.49}$ & $17\pm11$ & $<630$ & 
6.3 \\
\enddata
\tablenotetext{a}{Measured line energy in quasar rest frame, units
  eV.}
\tablenotetext{b}{Photon flux, units $\times10^{-5}$\,photons\,cm$^{-2}$\,s$^{-1}$}
\tablenotetext{c}{Equivalent width in quasar rest frame, units eV.}
\tablenotetext{d}{$1 \sigma$ velocity width, units km\,s$^{-1}$.}
\tablenotetext{e}{Improvement in C-statistic upon adding line to model.}
\tablenotetext{t}{Indicates parameter is tied.}
\label{emission-lines}
\end{deluxetable}

\clearpage

\begin{deluxetable}{lcccc}
\tabletypesize{\small}
\tablecaption{Warm Absorber Parameters from HETG Spectrum}
\tablewidth{0pt}
\tablehead{
\colhead{Parameter} & \colhead{Zone\,1} & \colhead{Zone\,2} & \colhead{Zone\,3} 
& \colhead{Zone\,4}} 
\startdata
$N_{\rm H}$$^{a}$ & $4.0^{+0.4}_{-0.4}$ & $2.1^{+0.4}_{-0.3}$ & $1.5^{+0.5}_{-0.4}$ & $10^{+6}_{-5}$\\
$\logxi$$^{b}$ & $-0.04^{+0.14}_{-0.18}$ & $1.07^{+0.11}_{-0.15}$ & $2.11^{+0.16}_{-0.11}$ & $3.7\pm0.4$\\
$v_{\rm out}$$^{c}$ & $-150^{+105}_{-100}$ & $+10^{+80}_{-90}$ & $+130^{+70}_{-60}$ & $<80$\\
$\Delta C$$^{d}$ & 946.7 & 130.3 & 46.0 & 10.3\\
\hline
\enddata
\tablenotetext{a}{Hydrogen column density, units $\times10^{21}$\,cm$^{-2}$.}
\tablenotetext{b}{Ionization parameter in log units.}
\tablenotetext{c}{Outflow velocity in units km\,s$^{-1}$. Positive values indicate inflow.} 
\tablenotetext{d}{Improvement in C-statistic upon the addition of the 
component to the model.} 
\label{absorbers}
\end{deluxetable}

\end{document}